\documentclass[lettersize,journal]{IEEEtran}
\usepackage{amsmath,amsfonts}
\usepackage[ruled,linesnumbered]{algorithm2e}
\usepackage{array}
\usepackage{textcomp}
\usepackage{stfloats}
\usepackage{url}
\usepackage{verbatim}
\usepackage{graphicx}
\usepackage{cite}
\usepackage{verbatim}
\usepackage{graphicx}
\usepackage{textcomp}
\usepackage{subfigure}
\usepackage{subfiles}  
\usepackage{mathrsfs}
\usepackage{xcolor}
\usepackage[numbers,sort&compress]{natbib}
\newcommand{\name}{WiIntruder~}
\newcommand{\sname}{WiIntruder}

\begin{document}

\title{Security Analysis of WiFi-based Sensing Systems: Threats from Perturbation Attacks
}

\author{Hangcheng Cao,~
        Wenbin Huang,~
        Guowen Xu,~
        Xianhao Chen,~
        Ziyang He,~
        Jingyang Hu,~
        Hongbo Jiang,~\IEEEmembership{Senior Member,~IEEE},
        and Yuguang Fang,~\IEEEmembership{Fellow,~IEEE}
}        

\maketitle

\begin{abstract}
Deep learning technologies are pivotal in enhancing the performance of WiFi-based wireless sensing systems. However, they are inherently vulnerable to adversarial perturbation attacks, and regrettably, there is lacking serious attention to this security issue within the WiFi sensing community. In this paper, we elaborate such an attack, called \sname, distinguishing itself with universality, robustness, and stealthiness, which serves as a catalyst to assess the security of existing WiFi-based sensing systems. This attack encompasses the following salient features: (1) Maximizing transferability by differentiating user-state-specific feature spaces across sensing models, leading to a universally effective perturbation attack applicable to common applications; (2) Addressing perturbation signal distortion caused by device synchronization and wireless propagation when critical parameters are optimized through a heuristic particle swarm-driven perturbation generation algorithm; and (3) Enhancing attack pattern diversity and stealthiness through random switching of perturbation surrogates generated by a generative adversarial network. Extensive experimental results confirm the practical threats of perturbation attacks to common WiFi-based services, including user authentication and respiratory monitoring.

\end{abstract}
\thispagestyle{empty}

\section{Introduction}
\label{sec:intro}
In WiFi networks, communication services need \textit{channel state information} (CSI)~\cite{ma2019wifi} to capture how wireless signals propagate in the physical space. It was originally used to assist data transmission, but its sequential patterns can also capture target object states~\cite{9796740,8613849,3485936,9141400,wang2016device}. In this context, WiFi has evolved into a potent sensing facilitator~\cite{wang2020learning,wang2018device}. Recent years have witnessed the emergence of a diverse array of innovative WiFi-based sensing applications~\cite{3326081,3467032,xiong2013arr}, such as activity recognition and vital sign monitoring. These WiFi-based sensing services have garnered intensive attention in both academia and industry, primarily owing to their non-intrusive nature and high privacy preservation. Moreover, deep learning models have gained significant attention in the quest to enhance sensing service performance. For example, EI~\cite{3241548} has devised a convolutional neural network to ensure the generalization of user activity features across distinct environments. Similarly, Widar3~\cite{3326081} has meticulously crafted a signal processing pipeline based on a hybrid convolutional-recurrent neural network, designed to distill environment-independent gesture features.

However, recent advances in security research~\cite{3423348, chc1, LiNPSKRS19} have revealed the potency of well-designed perturbations to mislead learning-based models, prompting them to yield erroneous results. While perturbation attacks are well-explored in fields like image classification~\cite{LiNPSKRS19} and voice recognition~\cite{3423348}, the WiFi sensing community has only recently begun to realize their relevance. Such a delay is concerning because sensing applications (e.g., vital sign monitoring and user authentication) have already been used for personal health monitoring and smart home, which will cause serious consequences~\cite{nan2023you}. For example, if perturbation attacks maliciously manipulate the respiratory rate monitoring service, it can trigger false alarms in healthcare systems and strain social medical resources. Unfortunately, existing perturbation mechanisms show limited practicality in real-world scenarios and thus result in users not being accurately aware of its dangers. For instance, Zhou et al.~\cite{3534618} target specific gesture recognition models with perturbation signals from customers' full-duplex devices, lacking universality. Similarly, recent works in \cite{liu2023exploring} and \cite{9796920} focus on specific learning-based models, neglecting universality across different sensing applications. Moreover, there are a few proposals for perturbation attacks toward WiFi communications~\cite{3484777,9609969,8792120} and other radio frequency-based sensing mechanisms (e.g., mmWave radar~\cite{yingyingchen}), but their effectiveness in attacking WiFi-based sensing remains uncertain. Therefore, one work of fully studying the practicability of perturbation attacks is still needed, which is critical for analyzing and assessing the security of WiFi sensing systems.

While perturbation attacks may seem straightforward, the practical implementation presents a significant challenge, as highlighted in previous works~\cite{ZhangZL0CZH21,WangHCLCW22,XieWKH22}. To be effective and practical, such attacks must simultaneously possess three essential features: i)~\textit{Universality}. The success of one perturbation attack hinges on its ability to affect a wide range of sensing models across different applications, without prior knowledge of the target model's structure, which will be practical in the sense that an attacker does not rely on too much knowledge to launch an attack. Therefore, unlike previous works~\cite{3534618,liu2023exploring,9796920}, one practical black-box attack should demonstrate its universality, i.e., enabling a single perturbation signal to mislead various sensing models. ii)~\textit{Robustness}. In real-world attack environments, perturbation signals are emitted from malicious devices and received by legitimate ones. During the propagation process, signals undergo inevitable distortion caused by device desynchronization (e.g., transmission time offset) and wireless channels (e.g., multi-path effect and energy attenuation). Thus, to make an attack effective, the attack strategy has to take the perturbation signal distortion into consideration. iii)~\textit{Stealthiness}. An effective perturbation attack must operate stealthily, remaining undetected by the target system, even in the absence of knowledge about the specific model structures.

\begin{figure}[t]
\centering
\includegraphics[width=0.5\textwidth]{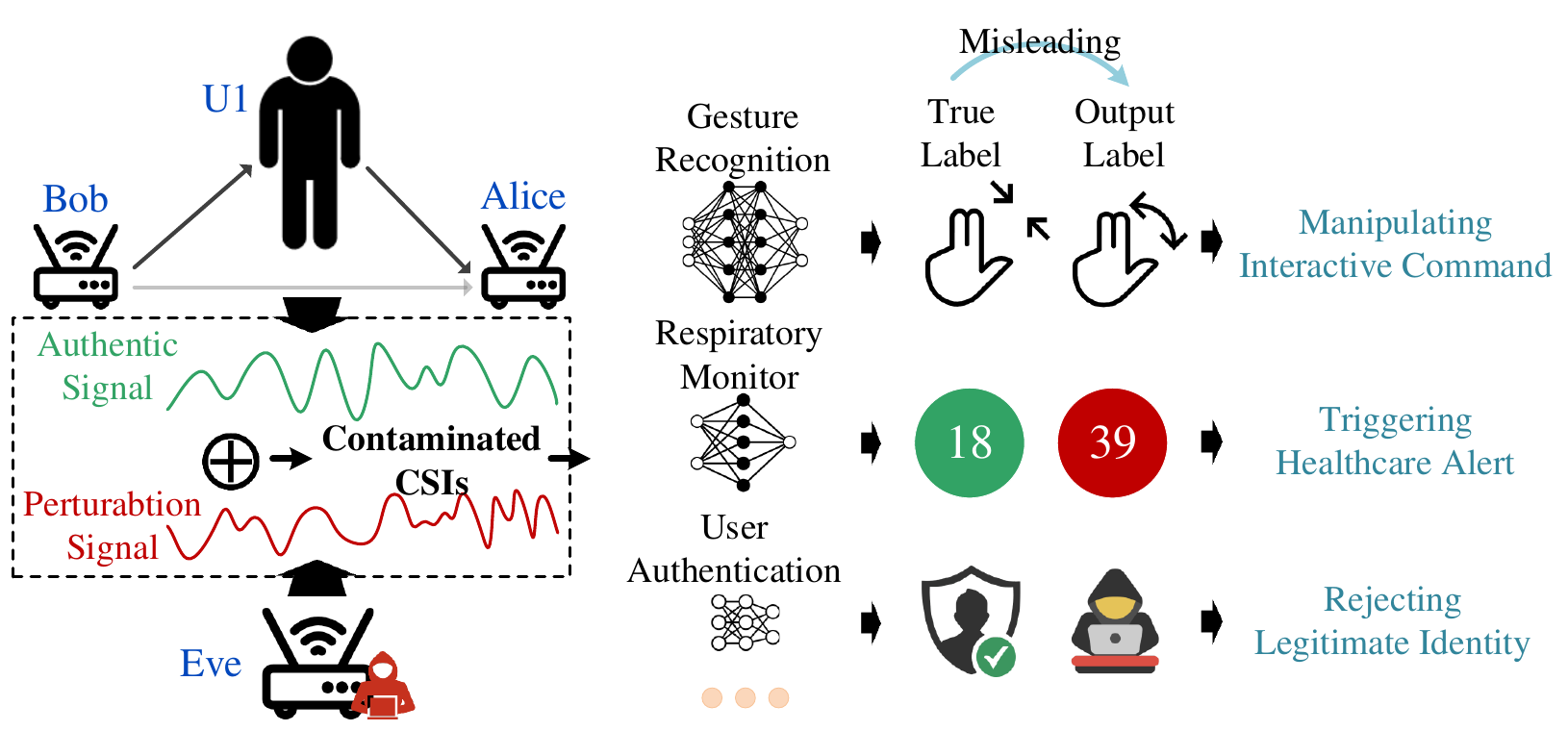}
\caption{Attack scenario of \sname: perturbation signals emitted from an attacker (Eve) to contaminate authentic ones (and hence CSIs) between one legitimate transmitter (Bob) to one receiver (Alice), thereby misleading sensing models to output false results.}
\label{fig:introScenario}
\end{figure}

In this paper, we design a practical black-box perturbation attack, \textbf{\sname}, possessing the above three merits, to reappraise the security of existing WiFi sensing systems. Its application scenario is illustrated in Fig.~\ref{fig:introScenario}. For WiFi-based sensing, CSIs represent wireless channel variations and hence channel state-specific information.
The attacker, known as Eve, generates perturbation signals and transmits them into the target physical space. Here, the legitimate receiver, Alice, receives contaminated wireless signals consisting of authentic signals from Bob and perturbed ones from Eve. In response to these perturbed signals, learning-based sensing models may make erroneous decisions, such as providing incorrect respiratory rates or rejecting valid user authentication attempts. The challenge lies in designing the perturbation signals while considering the practical aspects of attack implementation. 
First, variations in model structures and parameter settings across sensing applications create uncertainty in designing a universal perturbation signal. Second, in real-world attack scenarios, signal distortion during over-the-air propagation is inevitable, undermining attack performance. Lastly, while a universal attack supports the use of a single perturbation signal across models and applications, this static nature makes it vulnerable to existing defense mechanisms~\cite{3465397}. Thus, achieving a balance between the universality and stealthiness of these perturbations is of paramount importance.

To address the above challenges, we take the following steps to construct \sname: i) We have adopted the primary parameters, as CSI amplitude and phase in perturbation signal generation, to contaminate the original wireless channels across various applications consistently effective to attack the target system. Additionally, to guarantee the universality of the perturbation signal, we maximize the differences in user state-specific feature spaces across different models and applications.
ii) We analyze CSI amplitude and phase deviation introduced by signal distortion and then leverage a heuristic algorithm-driven perturbation generation mechanism by considering this deviation occurrence, to guarantee attack robustness. iii) We employ an energy-based generative adversarial network as our data augmentation pipeline to generate multiple surrogates of one perturbation and then dynamically switch the surrogates to launch more effective attacks. In this case, one perturbation noumenon corresponds to multiple attack patterns, to avoid then detection. After generating perturbation signals, we verify its intended impact on existing WiFi sensing applications. This result further demonstrates the importance of accurately assessing the threat of perturbation attacks, thereby providing users with secure sensing services. In a nutshell, we make the following major contributions:  
\begin{itemize}
\item{We propose an effective perturbation attack that simultaneously possesses universality, robustness, and stealthiness, thereby fully analyzing the security of WiFi-based wireless sensing systems.}

\item{We design a perturbation generation framework comprehensively handling interference factors in real-world attack implementation, e.g., model structure differences and signal distortion. }

\item{We conduct extensive experiments in four common sensing applications\footnote{Due to the limited space, we utilize four representative WiFi sensing applications~\cite{ma2019wifi,iu2014chann,572909} to verify the effectiveness of \sname, which covers four multiple scenarios named human-computer interaction, health monitoring, system security, and indoor localization.} and the results demonstrate that \name leads to a remarkable accuracy degradation, i.e., 79.6\% in activity recognition, 65.2\% in respiratory state monitoring, 68.3\% in user authentication, and 78.4\% in indoor localization with low probability of detection.}

\item{By verifying the performance of \sname, we further present the impact of perturbation attacks on the WiFi sensing applications, which can be used to assess security issues.}

\end{itemize}

The rest of this paper is organized as follows. Sec.~\ref{sec:relatedWork} presents the related works on WiFi sensing and perturbation attacks. Subsequently, Sec.~\ref{sec:background} describes the running mechanism of wireless sensing, attack model, perturbation attack definition, and feasibility analysis. System overview and technical modules are presented in Sec.~\ref{sec:sysDesign}, then the implementation and experimental results are reported in Sec.~\ref{sec:evaluation}. The related limitations and future works are reviewed in Sec.~\ref{sec:discussion}. Finally, we conclude our work in Sec.~\ref{sec:conclusion}.

\section{Related Work}
\label{sec:relatedWork}
In this section, we review existing WiFi sensing applications, then analyze the threat of perturbation attacks to them. 

\textit{WiFi-based sensing applications.}
WiFi sensing has spurred substantial efforts from both academia and industry to harness its potential in developing innovative applications. Existing research predominantly focuses on extracting CSI amplitude and phase variation patterns in both time and frequency domains to recognize user-specific states~\cite{6847948,3326081,3241548,chc,9941045}, like gestures and activities. For instance, Widar3~\cite{3326081} extracts domain-agnostic body-coordinate velocity profiles to facilitate cross-domain gesture recognition. In a similar vein, EI~\cite{3241548} employs an adversarial network-enabled framework to eliminate environmental information from raw CSIs, enabling cross-domain activity recognition. Furthermore, WiFi-based sensing has proven valuable in fall detection. WiFall~\cite{6847948} exploits the time variability and spatial diversity of CSIs to represent user behaviors, enhancing detection accuracy through a local outlier factor-based algorithm. WiFi sensing medium is also effective for respiration monitoring and indoor localization. MultiSense~\cite{3411816} treats respiration estimation as a blind source separation problem, enabling continuous monitoring of multiple individuals' respiration patterns. \cite{9380161} studies the corresponding relationship between received signal strength and user position to complete the localization task. To sum up, WiFi-based sensing applications continue to proliferate, offering users convenient and user-friendly services.

\textit{Perturbation attacks against WiFi sensing systems.}
The potential of deep learning technology gradually manifests in WiFi sensing~\cite{9796740,3326081,9141400,3534574}. Therefore, researchers have started paying close attention to the potential risks when wireless sensing systems face perturbation attacks. This attack leverages perturbation signals to contaminate original CSI patterns, thereby misleading learning-based models into making wrong decisions. WiAdv~\cite{3534618} launches an attack against one deep learning-powered gesture recognition system, which can mislead the system to output the targeted gesture label. Nevertheless, this work only presents the effectiveness of attacks against the specific model Widar3. To further enlarge the attack surface, Liu et al.~\cite{9796920} design a perturbation attack to manipulate CSI patterns of user behaviors, by jamming WiFi signals. This work only exhibits its threat on a single application, namely, behavior recognition, while IS-WARS~\cite{huang2021wars} endures a similar predicament. The recent work RAFA~\cite{liu2023exploring} designs a perturbation signal generation mechanism and considers interference factors generated in the real-world attack setting. However, it just presents the attack performance of an indoor localization system, thus the universality across different sensing models is unknown. The limitation of existing studies implies that it is important to explore the practicality of perturbation attacks, which can facilitate a deeper understanding of its hazard and assist in security analysis of WiFi sensing systems.

\textit{Perturbation attacks in other fields.}
Perturbation attacks receive intensive attention in other fields involving deep learning such as computer vision and speech recognition. Compared with WiFi sensing, image and audio are the more severely damaged sensing medium since their system significantly relies on deep learning techniques. Existing studies work well in the black-box setting and show the destructiveness of perturbation attacks on various applications, including 2D image classification~\cite{ChengDPSZ19}, semantic segmentation~\cite{yang2020adversarial}, and object detection~\cite{li2021playing}. One recent study~\cite{wei2020heuristic} also extends 2D image attacks to 3D video scenarios. Moreover, researchers also discover the threat of perturbation attacks imposed on intelligent voice systems. For example, AdvPulse~\cite{3423348} obtains the perturbation sample by a penalty-based universal generation algorithm while considering the over-the-air propagation time delay. VMask~\cite{zhaneprint} adds subtle perturbations to the recordings from a specific speaker and deceives the identity verification module to classify it as a legitimate identity. Nevertheless, compared with the WiFi sensing systems, they have different signal distortion and hardware imperfection levels. Therefore, we cannot directly adapt existing research results in other fields to WiFi systems. In this case, comprehensive exploration on the threat of practical perturbation attacks against WiFi-based sensing is highly interesting.

\section{Background and Preliminaries}
\label{sec:background}
In this section, we first present the background of WiFi sensing and our threat model. Then, we introduce the basic formulation of perturbation attacks and further highlight the feasibility of \sname's implementation.

\begin{figure*}[b]
\subfigure[Gesture recognition]{
\begin{minipage}[t]{0.23\linewidth}
\centering
\includegraphics[width=1\textwidth]{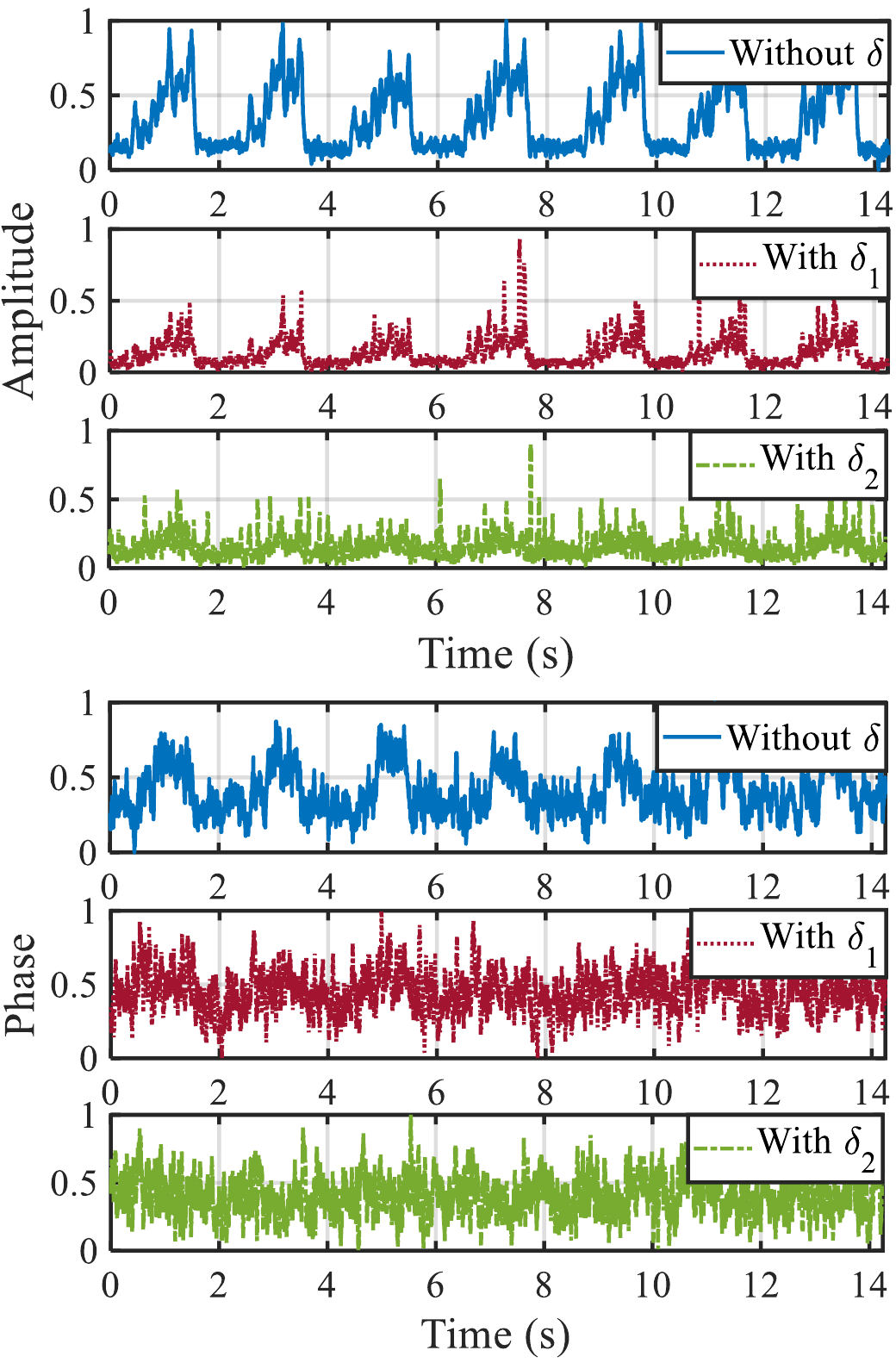}
\label{fig:gesture1}
\end{minipage}
}
\subfigure[Respiratory monitoring]{
\begin{minipage}[t]{0.23\linewidth}
\centering
\includegraphics[width=1\textwidth]{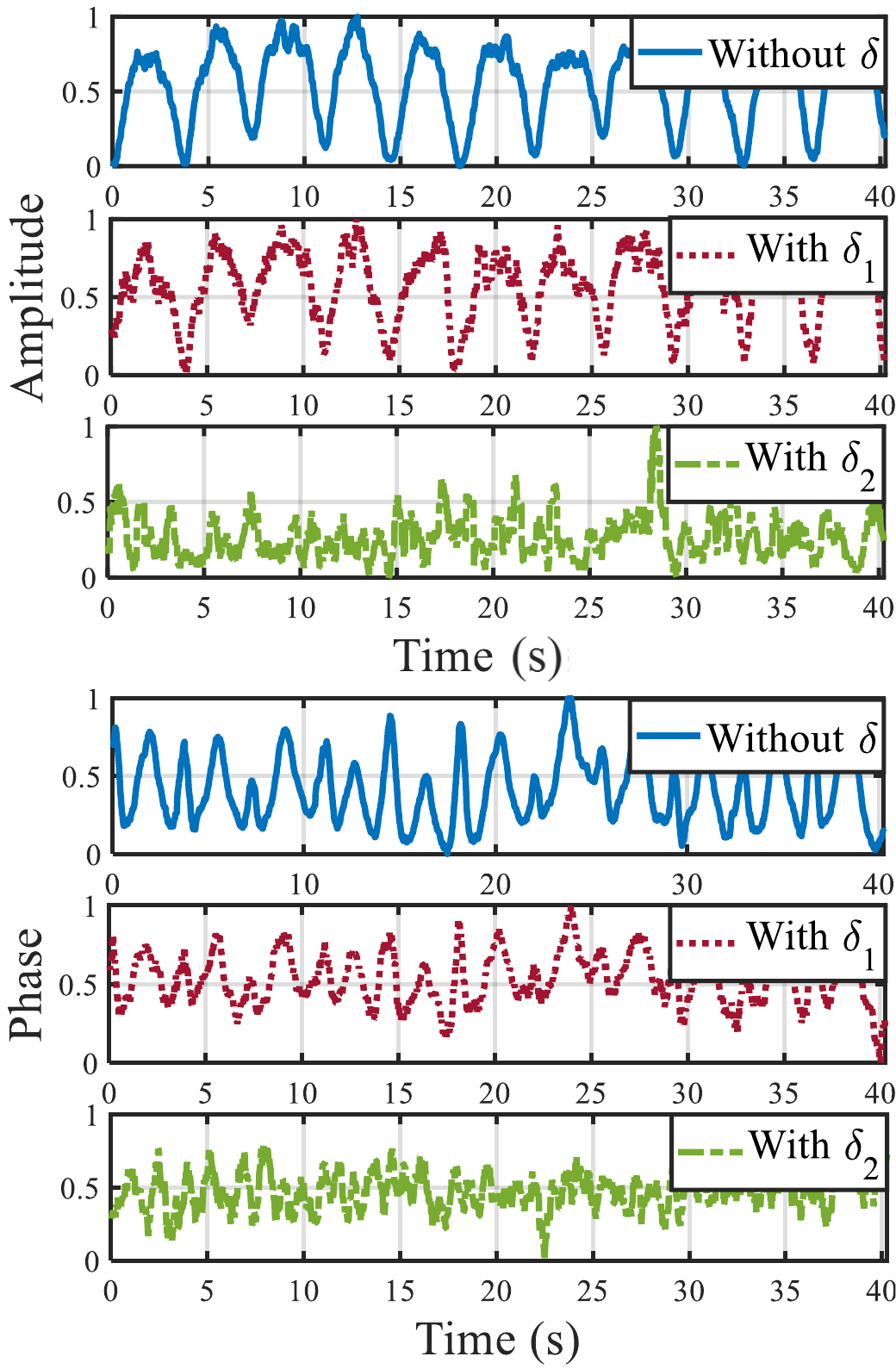}
\label{fig:respiratory1}
\end{minipage}
}
\subfigure[User authentication]{
\begin{minipage}[t]{0.23\linewidth}
\centering
\includegraphics[width=1\textwidth]{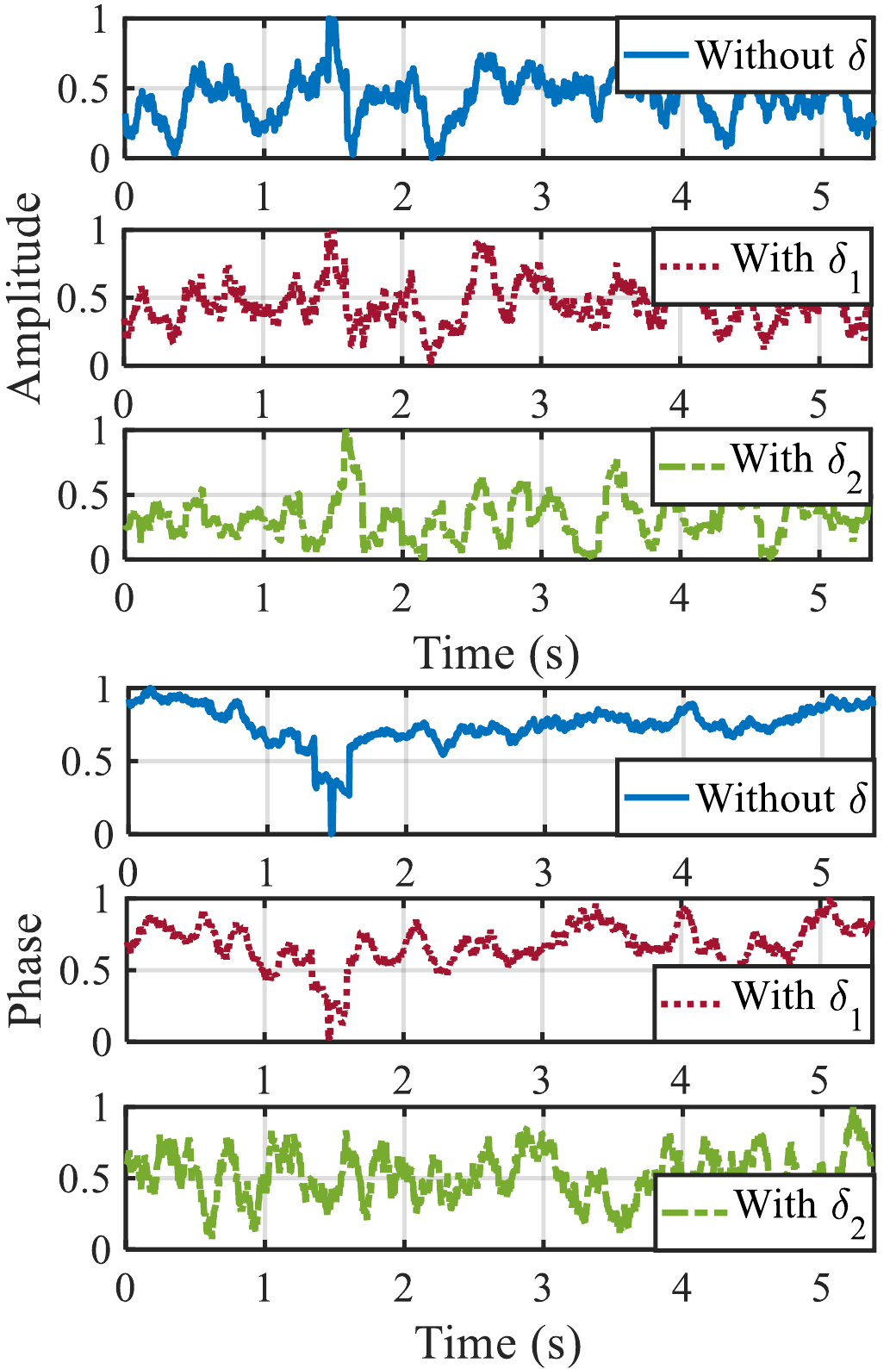}
\label{fig:authentication1}
\end{minipage}
}
\subfigure[Indoor localization]{
\begin{minipage}[t]{0.23\linewidth}
\centering
\includegraphics[width=1\textwidth]{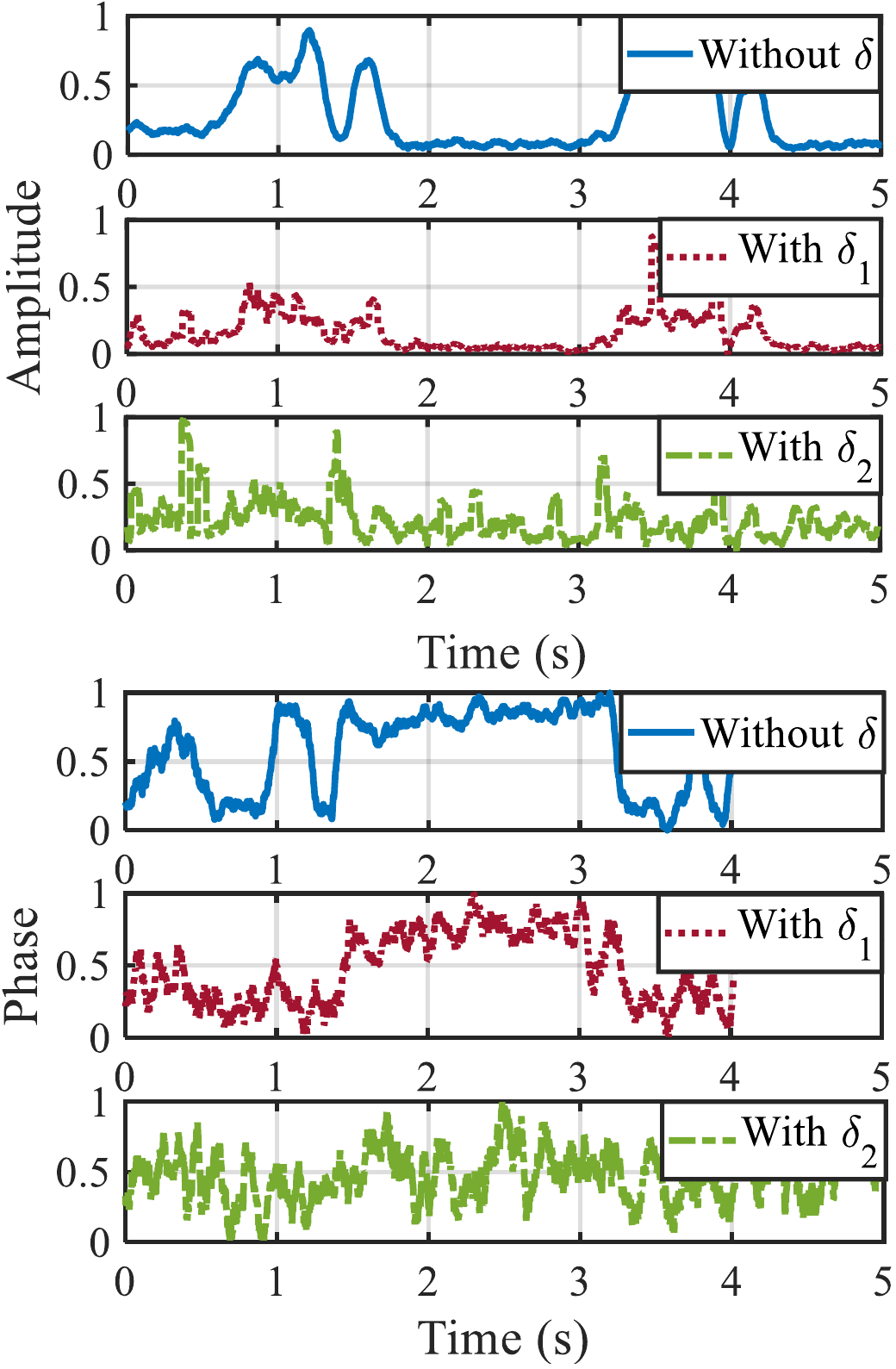}
\label{fig:indoor1}
\end{minipage}
}
\caption{Perturbation signals ${\delta_1}$ and ${\delta_2}$ utilized to contaminate authentic CSI amplitude and phase patterns in four common sensing applications, showing that ${\delta_2}$ offers the superior attack performance.}
\label{fig:feasiStudy1}
\end{figure*}

\subsection{WiFi-based Sensing Mechanism}
\label{subsec:wifisensing}
WiFi signals always experience wireless channel distortions during the propagation process, even one single moving object can change signal propagation paths and cause variations in WiFi channels (i.e., CSIs). Therefore, there is a correspondence between CSI patterns and object moving states. CSI is prerequisite for correct decoding in wireless communications and hence IEEE 802.11 protocol stack always utilizes \textit{long training sequence} (LTS)~\cite{iu2014chann} to estimate CSI. Especially, LTS is a public field in \textit{orthogonal frequency division multiplexing} (OFDM) preambles~\cite{cimysis}, thus any devices equipped with network interface cards can overhear LTSs and extract the corresponding wireless channel information. Giving the CSI ${h}_{n,m}$ of the $n$-th OFDM subcarrier in the $m$-th packet, when one WiFi transmitter sends LTS signal $s_{n,m}$, the received version is $y_{n,m} = h_{n,m} s_{n,m}$, where $h_{n,m}$ is the channel impulse response that can be modeled as:
\begin{align} \label{eq:csi_subc}
    {h}_{n,m} = \textstyle{\sum\limits_{l = 1}^{L}} \alpha_{n,m,l} e^{-j 2 \pi f_n \tau_{m,l}  } 
\end{align}
where $\alpha_{n,m,l}$ is the signal attenuation on the $l$-th signal propagation path, $f_n$ denotes the frequency of the $n$-th subcarrier, and $\tau_{m,l}$ is the time delay consisting of the static paths from surrounding reflectors and the dynamic ones caused by the moving object. From Eqn.~(\ref{eq:csi_subc}), once there is an object in moving states, the dynamic path will constantly cause changes in CSI patterns. Therefore, CSI amplitudes and phases can effectively capture the corresponding user states and act as the raw characterization of WiFi sensing systems. 

\subsection{Threat Model}
\begin{itemize}
\item{\textbf{Attack scenario.}} In line with existing adversarial perturbation attacks, the goal of \name is to bring down the performance of wireless sensing systems. 
However, our study differentiates itself by striving to employ a single perturbation signal capable of simultaneously attacking various models and applications. The attack scenario is visually represented in Fig.~\ref{fig:introScenario}, which comprises four key roles: Eve (an attacker), Alice (an AP or base station), Bob (a sensing device), and U1 (a legitimate user). Let us consider a common scenario: U1 is the object who performs movements (e.g., gestures and walking) in the vicinity of Bob and Alice, causing wireless channel changes. Bob sends LTS $s_{n,m}$ to Alice, who estimates CSI $h_{n,m}$ and hence user states using the received $y_{n,m}$ as described in Eqn.~\eqref{eq:csi_subc}. Meanwhile, Eve strategically emits well-designed perturbations to taint the user state-specific $h_{n,m}$ decoded by Alice, interference with normal sensing services.

\item{\textbf{Attacker ability.}}
With the attack setting of existing works~\cite{liu2023exploring,3534618,9796920}, we enable the following capabilities of Eve: i) For better concealment, Eve's transmitter should be equipped with the antenna quantity no larger than the common WiFi COTS and \name utilizes one antenna for emitting perturbations; ii) Due to the concealment, Eve can be located at any unobtrusive position in the target environment to launch attacks; iii) Eve possesses the full knowledge of sensing models utilized for constructing the perturbation generation mechanism and then launches attacks against new ones obeying a black-box manner; iv) To maintain practicality, Eve operates without the need of information coordination with Alice/Bob, such as informing them of signal transmission times and wireless channel states;
v) Eve can leverage publicly available operational information about communication protocols and hardware facilities to gain insights into the operations of Alice and Bob, including details like carrier frequency bands and encoding methods; and vi) We assume that Alice and Bob have no customized shielding settings against electromagnetic interference. The rationality of this assumption lies in that this shielding is rarely deployed in daily life, since it interferes with communication services and carries additional costs.
\end{itemize}

\subsection{Adversarial Perturbation Attack}
In this subsection, we elaborate the idea of our perturbation attack design. Given a learning-based sensing model $\mathcal{M}( \cdot )$, the user state ${\gamma_k}$ mapping to input ${H_k}$ (the $k$-th CSI frame segmented from $[h_{n,m}]$) can be inferred. In our scenario, Bob sends LTS $s_{n,m}$ with CSI of $h_{n,m}$, and Eve emits the perturbation signal $\delta_{n,m}$ (i.e, $h^{\delta}_{n,m}s_{n,m}$) with the channel state $h^{Eve}_{n,m}$ (caused by wireless propagation and device desynchronization as described in Sec.~\ref{subsec:roubust}) between Eve and Alice. Therefore, Alice obtains the contaminated signal $h_{n,m}s_{n,m}+h^{Eve}_{n,m}h^{\delta}_{n,m}s_{n,m}$ and the detected CSI is described as $h_{n,m}+h^{Eve}_{n,m}h^{\delta}_{n,m}$ (abbreviated as $H_k+H_{Eve}H_{\delta}$). \name aims to mislead the model to output false results, i.e., $\mathcal{M}({H_k} + H_{Eve}{H_{\delta}})\ne{\gamma_k}$. To achieve this goal, an attacker designs $\delta$ by solving the following maximization problem:
\begin{equation}\label{eqn:goal1}
\mathop {\arg \max }\limits_{{H_\delta}} \Gamma (\mathcal{M}(H_k + H_{Eve}{H_{\delta}}),{\gamma_k}){~s}{.t}{\rm{. }}\left\| {{H_\delta}} \right\|_{\infty} < \varepsilon
\end{equation}
where $\vert\cdot\vert_{\infty}$ denotes the $\infty$-norm and the loss function $\Gamma(\cdot)$ denotes the distance metric capturing the difference between the authentic and misleading results. To avoid the detection of the perturbed signal, a small constant $\varepsilon$ is considered to limit the $\delta$-induced wireless channel variation, hence lower the probability of being detected. In the optimization process, \name aims to maximize the difference between contaminated and authentic CSI patterns to effectively perturb user state-specific information while maintaining low detection.

\subsection{Feasibility Analysis}
\label{subsec:feasibility}
The prerequisite for implementing \name is the universality of perturbation signals, meaning that one perturbation can contaminate authentic CSIs and consequently affect the features of multiple sensing models simultaneously. However, different sensing models utilize distinct high-level features. For instance, Widar3~\cite{3326081} relies on body-coordinate velocity profiles, while SLNet~\cite{yang2023slnet} employs learning-enhanced spectrograms. The question arises: can Eve employ a single perturbation signal to contaminate diverse high-level features and mislead corresponding learning-based models?~This question still remains open, which is the focus of this paper. Fortunately, as Eqn.~(\ref{eq:csi_subc}) illustrates, these high-level features are derived from two primary CSI parameters, namely amplitude and phase. Under these circumstances, \name exhibits the potential to launch universal attacks against various sensing models by manipulating only these two parameters. Subsequently, we will conduct a preliminary experiment to verify the feasibility of the aforementioned analysis.

\begin{figure}[t]
\subfigure[Body-coordinate velocity profile]{
\begin{minipage}[t]{0.46\linewidth}
\centering
\includegraphics[width=1\textwidth]{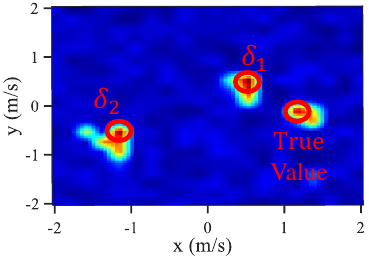}
\label{subfig:bvp2}
\end{minipage}
}
\subfigure[Learning-enhanced spectrogram]{
\begin{minipage}[t]{0.46\linewidth}
\centering
\includegraphics[width=1\textwidth]{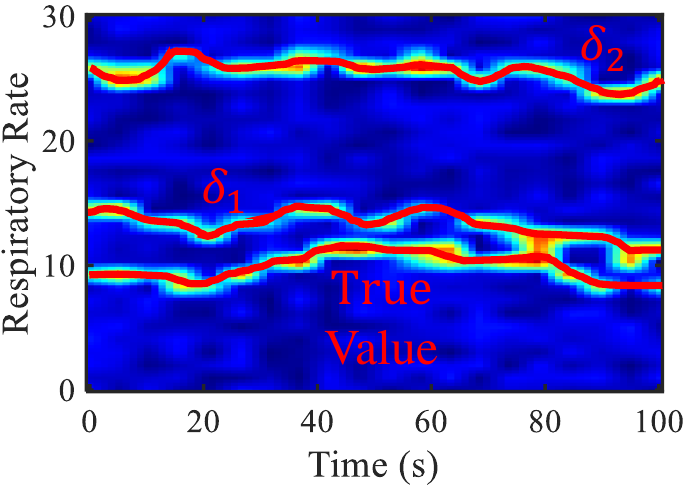}
\label{fig:respiratory2}
\end{minipage}
}
\subfigure[Time-frequency spectrogram]{
\begin{minipage}[t]{0.46\linewidth}
\centering
\includegraphics[width=1\textwidth]{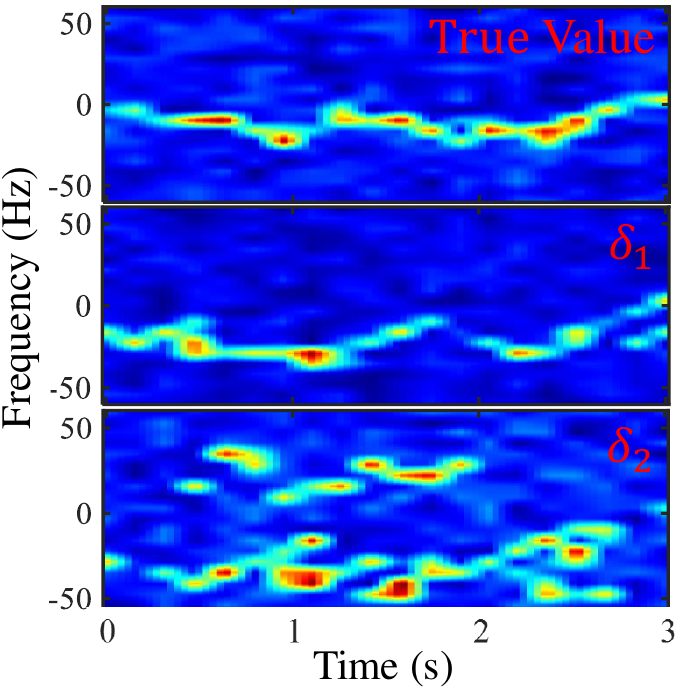}
\label{subfig:authentication2}
\end{minipage}
}
\subfigure[Learning-based latent features]{
\begin{minipage}[t]{0.46\linewidth}
\centering
\includegraphics[width=1\textwidth]{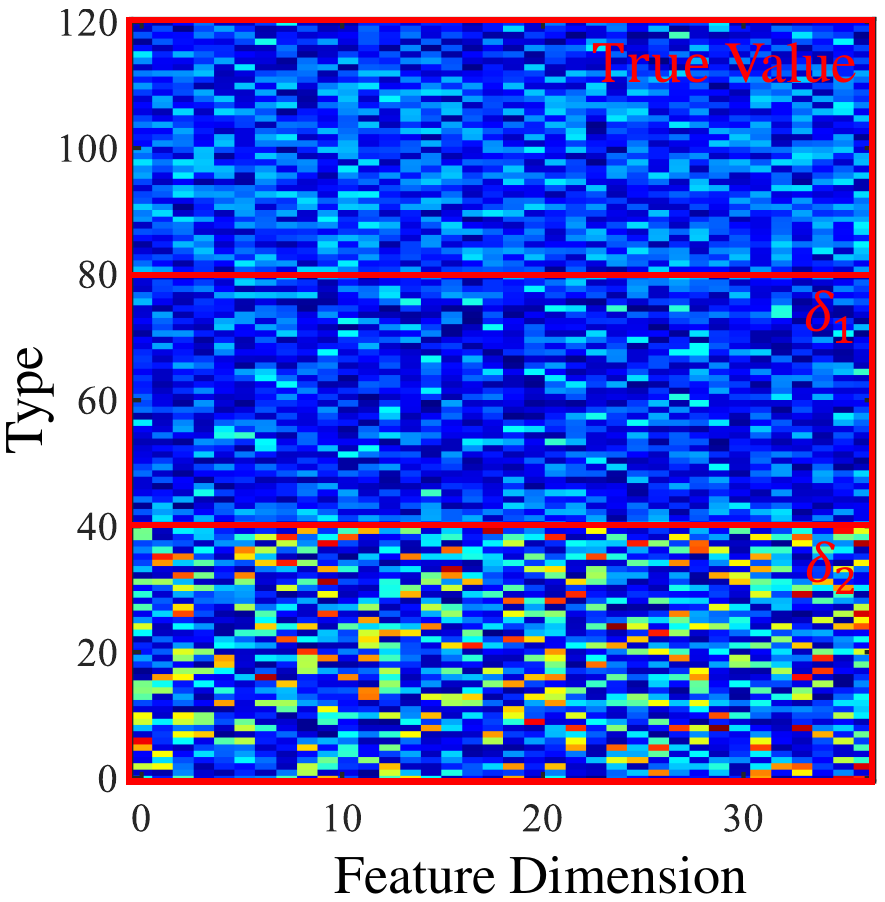}
\label{subfig:localization2}
\end{minipage}
}
\caption{High-level features and their containmated versions corresponding to the random perturbation ${\delta_1}$ and the optimized one ${\delta_2}$, in (a) gesture recognition, (b) respiratory monitoring, (c) user authentication, and (d) indoor localization.}
\label{fig:feasiStudy2}
\end{figure}

There are four learning-based models involved in this experiment, namely, Widar3~\cite{3326081} for gesture recognition, SLNet~\cite{yang2023slnet} for respiratory rate monitoring, MultiAuth~\cite{3467032} for user authentication, and DAFI~\cite{3494954} for indoor localization. Relative experiment settings are described in Sec.~\ref{subsec:experimentSet} in detail. There are two types of perturbation signals utilized to contaminate authentic CSI patterns of the aforementioned sensing applications: ${\delta_1}$ is randomly generated and ${\delta_2}$ is the optimization version of ${\delta_1}$ provided by \name (the detail is given in Sec.~\ref{sec:sysDesign}). Fig.~\ref{fig:feasiStudy1} presents the CSI amplitude and phase patterns with and without perturbation signals. Authentic CSI envelopes are severely contaminated by ${\delta_2}$ and the performance of the random perturbation ${\delta_1}$ is relatively weaker. Let us take gesture recognition (performing one gesture seven times) as an example in Fig.~\ref{fig:gesture1}. For example, each CSI segment owns one main crest with specific slopes, but the contaminated one of ${\delta_2}$ does not exhibit similar traits. Furthermore, the high-level features utilized in the four applications with and without perturbation are presented in Fig.~\ref{fig:feasiStudy2}. This result shows the following key points: ${\delta_2}$ maintains the oustanding contaminating ability in high-level features, e.g., taking the body-coordinate velocity profile information away from the original value. Due to lacking the optimization process, ${\delta_1}$ exhibits limited threat, i.e., the contaminated raw and high-level features still present similar patterns in CSI envelope shape, amplitude, and value distribution with the authentic one.

\noindent\textbf{Summary.}
Preliminary experiments demonstrate that one carefully designed single perturbation signal has the ability to perform universal attacks across various wireless sensing applications and models. However, to ensure its performance in the real-world setting, \name should design an effective perturbation generation and optimization mechanism to ensure its practicality.

\begin{figure}[b]
\centering
\includegraphics[width=0.5\textwidth]{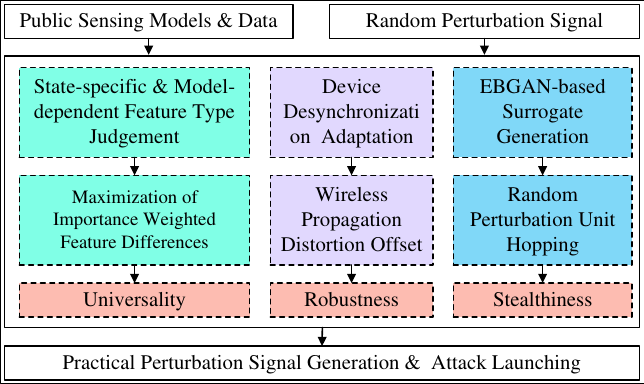}
\caption{The archiectural overview of \sname, showing the process of utilizing public sensing models and data as the raw material to remould random perturbations with three merits to launch practical attacks.}
\label{fig:overview}
\end{figure}

\section{System Design}
\label{sec:sysDesign}
This section first overviews the workflow of \name and then describes its technical modules on how to generate universal, robust, and stealthy perturbation signals to launch practical attacks.

\subsection{Workflow Overview}
As shown in Fig.~\ref{fig:overview}, the workflow consists of three main parts that endow random perturbation signals with three merits, namely, robustness, universality, and stealthiness. After obtaining public sensing models and data, the first step of \name is to judge feature types (i.e., state-specific and model-dependent) and maximize the difference between authentic and contaminated features across models to improve the transferability of perturbation. Subsequently, the distortion (caused by device desynchronization and wireless propagation) between the emitted and received perturbation signals is seriously handled by treating them as optimization factors in the perturbation generation process. The final step is to output multiple surrogates of one perturbation by an \textit{energy-based generative adversarial network} (EBGAN) model~\cite{zhao2016energy}, thereby keeping the diversity of perturbation patterns and hence ensuring the attack's stealthiness. Relying on the above steps, random perturbations are converted into optimized ones to launch practical attacks.

\begin{figure}[t]
\centering
\includegraphics[width=0.46\textwidth]{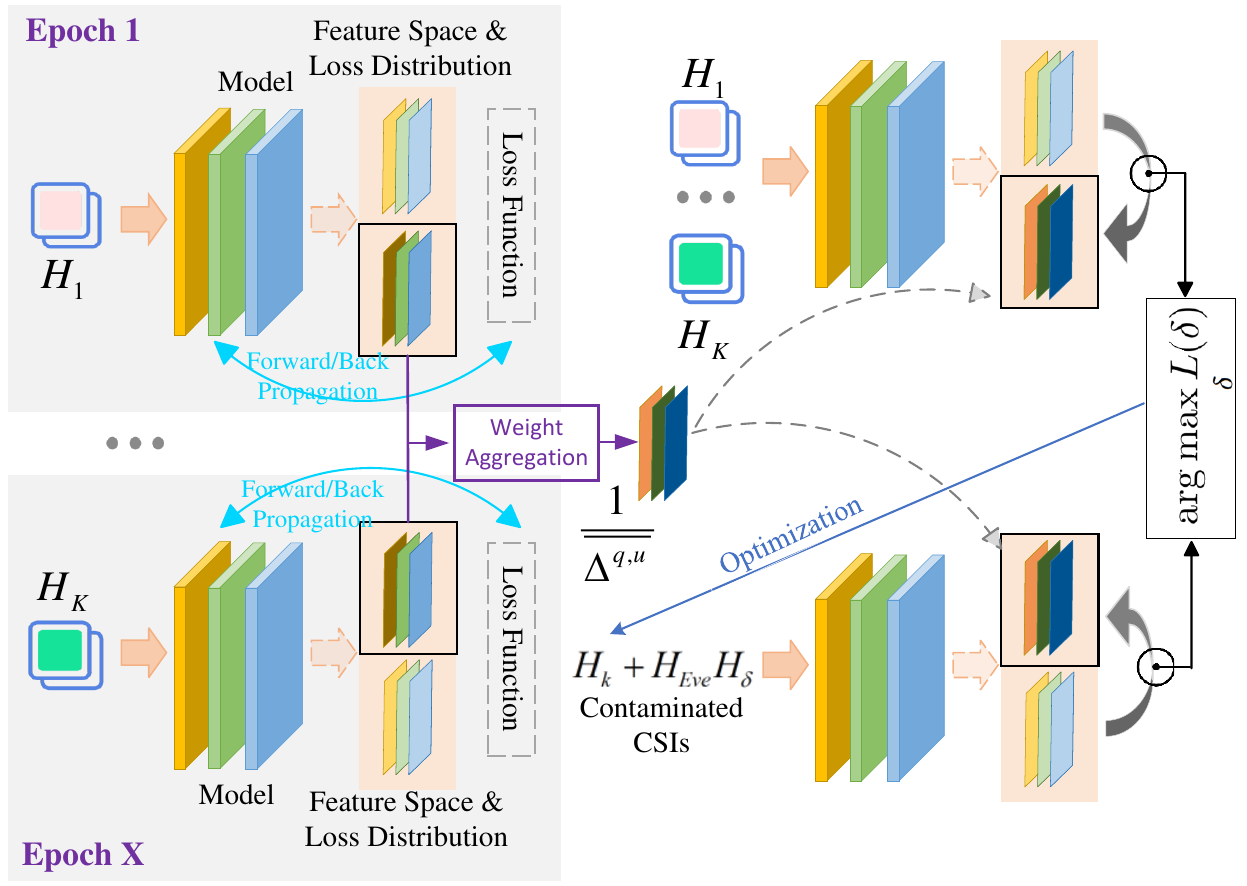}
\caption{Illustration of importance weight-driven perturbation optimization.}
\label{fig:crossModel1}
\end{figure}

\subsection{Universal Perturbation}
\label{subsec:uni}
Attack universality depends on whether a perturbation holds the transferability property, that is, the designed $\delta$ is applicable to effectively contaminate CSIs and hence the corresponding raw/high-level signal features across sensing models. However, existing works expose weak transferability because of indiscriminately disrupting the whole feature space of all layers in the perturbation optimization process~\cite{salzmann2021learning,huang2019enhancing}. Actually, the feature space should be categorized as two types, namely, state-specific and model-dependent, referring to their contributions to decision making~\cite{wang2021feature}. The former type carries critical information representing user states and thus can dominate outputs, while the latter just embeds the specific information of a model. Therefore, emphatically disrupting the user-state sub-component can further affect all model decisions and enhance the transferability of adversarial perturbations. Nevertheless, model-dependent features as the appendant are also generated in the model training and parameter optimization process, which hinders transferability. In this case, discriminately disrupting the two sub-components to avoid falling into model-dependent local optimization is critical in improving the transferability. 

Thus, the key question here is how to discriminately treat the two sub-components in the perturbation generation/optimization process. Based on the above analysis, the state-specific features contribute more to the model decision making and they correspondingly own large gradients in the model training phase. Furthermore, gradient value calculation relies on each feature element's loss, thus the loss $\Delta_{H_k}^{q,u}$ of the $q$-th model's $u$-th layer element can represent the corresponding feature contribution. To constantly evaluate the feature importance (contribution), we aggregate loss $\Delta_{H_k}^{q,u}$ of all training epochs to calculate the final weight. Fig.~\ref{fig:crossModel1} shows the importance weighted perturbation generation. We take the gesture recognition as an example to describe this process. Each batch consists of CSI samples of multiple gestures, while the gesture label represents one user state. When inputting samples from distinct gestures, the feature space of state-specific sub-components will change due to their own high losses, but the variations of the model-dependent ones are relatively small. After obtaining the aggregated loss $\overline{\Delta^{q,u}}$, \name aims to maximally disrupt the state-specific feature space and lower the impact on the model-dependent one. Therefore, we retune the loss function to disrupt the two-typed features by assigning them with different weights based on $\overline{\Delta^{q,u}}$ as follows:
\begin{footnotesize}
\begin{equation}\label{eqn:gradient2}
 \mathcal{L}(\delta) = {\sum\limits_{q=k=u=1}^{\rm{Q,K,U}} \frac{1}{{\overline {\Delta^{q,u}} }}} {\odot (|{\mathcal{M}^{q,u}}(H_k+{H_{Eve}{H_{\delta}}}) - {\mathcal{M}^{q,u}{(H_k)}|)}} 
\end{equation}
\end{footnotesize}
where ${\mathcal{M}^{q,u}}(H)$ represents the feature space of the $u$-th layer of the $q$-th sensing model, when inputting $H$, and $\odot$ denotes the Hadamard product. $H_{Eve}$ depends on the wireless channel between Alice and Eve, while the way of obtaining it is described in Sec.~\ref{subsec:roubust}. State-specific features own relatively higher values in $\overline {\Delta^{q,u}}$, which represents their importance on model decision making, and thus \name needs to put focuses on how to disrupt them. After obtaining the importance weight, we aim to maximize the difference between the authentic feature space and the contaminated one, by assigning the state-specific features with low weights of $\frac{1}{\overline {\Delta^{q,u}}}$. In this setting, Eqn.~(\ref{eqn:gradient2}) focuses on enlarging the difference in the low-weight (i.e., state-specific) feature space. Finally, we take Eqn.~(\ref{eqn:gradient2}) into Eqn.~(\ref{eqn:goal1}) and formulate the optimization as follows: 
\begin{equation}\label{eqn:gradient3}
\arg \mathop {\max L(\delta)}\limits_\delta{~s}{.t}{\rm{. }}\left\| {{H_\delta}} \right\| < \varepsilon 
\end{equation}

\begin{figure}[t]
\subfigure[Difference distribution]{
\begin{minipage}[t]{0.46\linewidth}
\centering
\includegraphics[width=1\textwidth]{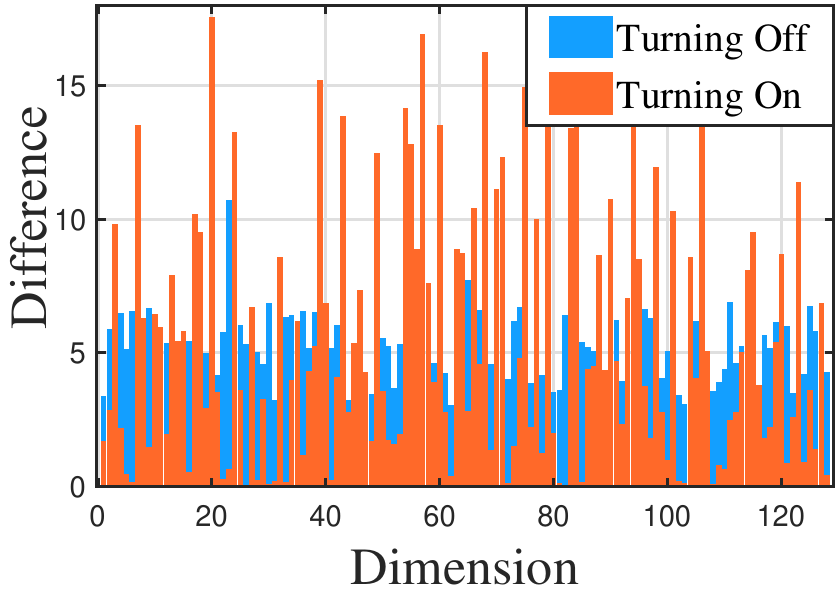}
\label{fig:gradientDis}
\end{minipage}
}
\subfigure[Recognition accuracy]{
\begin{minipage}[t]{0.47\linewidth}
\centering
\includegraphics[width=1\textwidth]{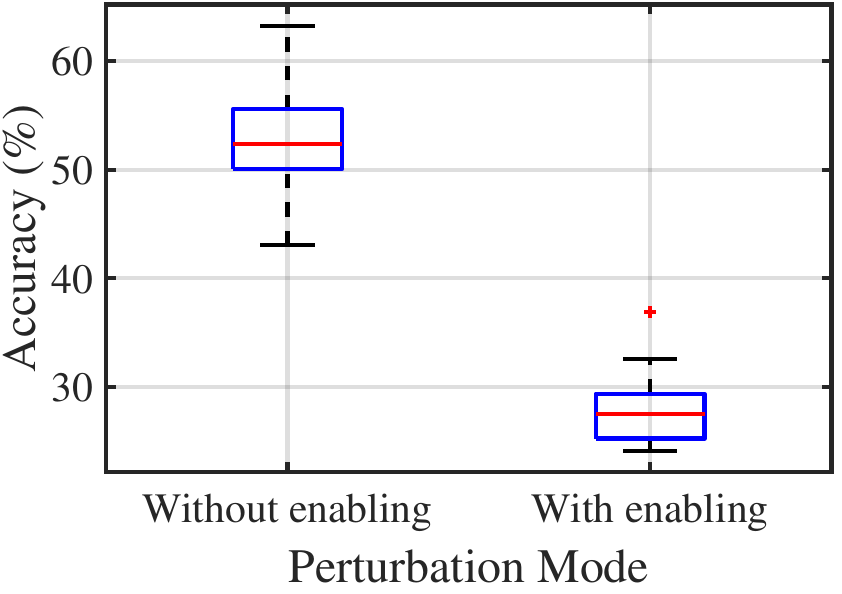}
\label{fig:featureDiff}
\end{minipage}
}
\caption{Feature difference of the recurrent neural network layer in Widar3 shown in (a), and the performance with/without turning on perturbation module presented in (b).}
\label{fig:corssModel2}
\end{figure}

To demonstrate the effectiveness of our weight-driven cross-model perturbation mechanism, we record the difference distribution in the 128-dimensional recurrent neural network layer of Widar3 under on-off perturbation cases. Fig.~\ref{fig:gradientDis} displays the difference (Euclidean distance) distribution of two cases, and their variance is 2.37 and 22.59, respectively. The key observation is that \name drives the difference to a few user-specific elements for discriminately treating distinct feature types. Subsequently, we compare the gesture recognition performance variation with the one enabling the weight-driven perturbation module. As shown in Fig.~\ref{fig:featureDiff}, our module leads to an obvious performance drop off of over 40\%. To sum up, by discriminatively treating two-typed features using the importance weight-driven difference maximization mechanism, the transferability of perturbations is further improved.

\subsection{Robust Perturbation}
\label{subsec:roubust}
Perturbation signals from generation to coming into force experience one critical stage named over-the-air propagation. This stage inevitably causes new distortions that, if not properly handled, inevitably affect attack practicality. The following content details two main distortions.

\textit{Device desynchronization distortion.} 
Clock desynchronization between Bob and Eve hinders the alignment of perturbation and authentic data signals. In real-world attack settings, Eve emits perturbations that propagate through wireless channels, and then Alice receives the mix of signals transmitted from Bob and Eve. \name expects the mixed (contaminated) CSIs to effectively mislead the decision making of sensing models. This expectation requires Eve to know the precise data emission time of Bob for signal alignment. Obviously, Bob and Eve cannot collaborate with each other for clock synchronization. One may utilize the preamble to detect Bob's data packet transmission, but it costs considerable time for signal processing operations and thus still cannot meet the fine-grained synchronization~\cite{nagantial}. In this case, \name overcomes this issue by enabling the robustness of perturbation signals when facing clock desynchronization, which brings a random delay $\Delta t$ between the two devices and can be expressed as a phase offset $e^{-j2\pi \Delta t{f_n}}$ of the $n$-th subcarrier. Moreover, we know that the oscillators of Alice and Eve always differ, thus causing imperfect signal processing and introducing extra random offsets~\cite{zhu2017calibrating}, namely, \textit{carrier frequency offset} (CFO), \textit{sampling frequency offset} (SFO), \textit{packet detection delay} (PDD), and \textit{carrier phase offset} (CPO). CFO is caused by the residue error in a phase-locked loop and it can be denoted as $e^{-j2\pi f_{\Delta c}}$, where $f_{\Delta c}$ is the carrier frequency difference. Since SFO and PDD impose the same phase impact and thus their offsets can be uniformly written as $e^{-j2\pi {f_b}n\tau _{sp}}$, where $f_b$ is the bandwidth between two adjacent subcarriers and $\tau_{sp}$ is the phase error. Nevertheless, current WiFi-based sensing mechanisms have proposed a series of solutions to handle these offsets such as linear fit~\cite{kotaru2015spotfi} and multi-antenna collaboration~\cite{ZengL0LW021}. CPO is a fixed value and can be readily removed in the hardware initialization phase. Therefore, the only device desynchronization factor that should be considered by Eve is $e^{-j2\pi \Delta t{f_n}}$ that imposes a fixed phase offset across data packets. Finally, we renovate the universal perturbation as ${H_\delta}[n,:]{e^{-j2\pi \Delta t{f_n}}}$. Hereby, we add one random phase offset selected from $[0, 2\pi]$ to all packets corresponding to one perturbation, to represent the impact of time delay $\Delta t$.

\textit{Wireless propagation distortion.} 
Perturbation signals propagating in the physical space between Eve and Alice experience wireless channel-dependent distortion. Taking transmitting a data packet as an example, Eve with an antenna obtains one $N \times {N_{ant}}$ wireless channel matrix, where $N_{ant}$ is the number of Alice's antennas. When emitting the perturbation $\delta$, the CSI becomes $H_{eve}{H_{\delta}}$, with consideration of the wireless channel state $H_{eve}$ between Eve and Alice. Therefore, if Eve wants to ensure the perturbation signal received by Alice as expected, obtaining $H_{eve}$ to precode emitted data in advance is necessary. For legitimate clients, Alice leverages LTS to readily estimate CSI as described in Sec.~\ref{subsec:wifisensing}. However, this approach is not feasible for Eve as it does not have any coordination with Alice. Fortunately, WiFi beacon protocol and channel reciprocity provide another way to obtain $H_{eve}$. Beacon frames are periodically sent out by one AP at specific time intervals to inform the potential clients Alice and Eve about the existence of the wireless network. Eve further obtains information such as Alice's name and encryption method from the beacon frame to initiate a network connection. During this interaction process, NICs can sniff and decode CSI information between Alice and Eve, such as Nexmon and ESP32 CSI tool used for 802.11a/g/n/ac~\cite{9673102,80211}. Intel has also developed new CSI reporting features for the latest Wi-Fi chipsets, such as the Wireless-AC9260/9560 and Wi-Fi 6 AX200/201 series~\cite{9673102}. After obtaining CSIs from Alice to Eve or in the opposite direction, one can infer the channel through channel reciprocity. Therefore, the bidirectional channels between Alice and Eve are equal. Relying on the above argument, Eve can obtain the channel state information $H_{eve}$ and process the distortion part as $H_{eve}[n,:]{e^{ - j2\pi \Delta t{f_n}}}$ (abbreviated to $H_{Eve}$), from which the perturbation signal can be designed with choice of ${H_\delta}[n,:]$.

\subsection{Perturbation Signal Generation}
\label{subsec:signalGen}
The essence of perturbation generation is to optimize the element of $\delta$ (and hence $H_{\delta}$) to maximize Eqn.~(\ref{eqn:gradient3}), to meet the demand of robustness and universality. Considering the fast convergence and easy adaption of \textit{particle swarm optimization} (PSO) method~\cite{ebericle}, we leverage it to optimize perturbations. \name utilizes one perturbation to contaminate multiple successive data packets, thus the number of elements that should be optimized is $N$ (being equal to the number of subcarriers). That is there are $N$ particle positions that should be searched in the value range of $[-\varepsilon,\varepsilon]$. \name leverages the objective function in Eqn.~(\ref{eqn:gradient3}) to play the role of the fitness function $\psi$, which guides the optimization direction of particle positions. In the initialization phase, the positions of $P$ particles in one swarm $Z$ are randomly selected in the searching space. In the $r$-th iteration, $Z$ is denoted as $\{ \overrightarrow {z_1^r} ,\overrightarrow {z_2^r} ,...,\overrightarrow {z_P^r} \}$. The $p$-th particle utilizes one $N$-dimensional vector to represent the position $[{z_{p,1}^r}, {z_{p,2}^r} ...,{z_{p,N}^r}]$. The particle $\overrightarrow {z_p^r}$ then computes current velocity $\overrightarrow {v_p^r}  = [{v_{p,1}^r} ,{v_{p,2}^r} ...,v{_{p,N}^r} ]$ as follows: 
\begin{small} 
\begin{equation}\label{eqn:pso1}
v_{_{p,n}}^r = w * v_{_{p,n}}^{r - 1} + {c_1} * {b_1}({d_{p,n}} - z_{p,n}^{r - 1}) + {c_2} * {b_2}({d_{glo,n}} - z_{p,n}^{r - 1})
\end{equation}
\end{small}

\begin{equation}\label{eqn:pso2}
z_{_{p,n}}^r = z_{_{p,n}}^{r - 1} + v_{_{p,n}}^r
\end{equation}
where $d_{p,n}$ represents the best position of the $p$-th particle, and ${d_{glo,n}}$ is the best position in the whole swarm. Each particle jointly utilizes the local $d_{p,n}$ and social ${d_{glo,n}}$ information to update and optimize its velocity and position, respectively. ${c_1}$ and ${c_2}$ are learning factors to weight the importance of personal and social parts in the position updating process, while both are chosen to be 0.5 for our study. ${b_1}$ and ${b_2}$ are random values selected in $[0,1]$, to ensure the diversity of position search and updating. $w$ is the inertia weight that decides the impact of the previous velocity on the current iteration. Its initial value is 1, with an attenuation coefficient 0.9. In each position update, \name checks whether the current perturbation meets the $\infty$-norm bound or not. The above perturbation optimization process is summarized in Algorithm~\ref{psoAlg}. After obtaining the near-optimal perturbation set $Z$, we further evaluate its robustness by adapting to the phase and amplitude variations caused by device desynchronization and wireless propagation distortion (as described in Sec.~\ref{subsec:roubust}), and finally retain the best-qualified one.

\begin{algorithm}
\caption{PSO-driven Perturbation Generation}\label{psoAlg}
\KwData{sensing model $\mathcal{M}^q$, CSI samples $H_k$, weight $\frac{1}{\overline {\Delta^{q,u}}}$, norm bound $\varepsilon$, PSO parameters $\{w,{c_{1,2}},{b_{1,2}},P,R\}$, fitting function $\psi$}
\KwResult{optimized particle (perturbation) set $Z$}
$Z \gets$ randomly sampling $P$ particles

\For{$p < P + 1$}{
\tcp{\footnotesize{checking the norm bound}}
$\theta _p \gets \psi_{\mathcal{M}^{q,u}, H_k,\frac{1}{\overline {\Delta^{q,u}}}}(\overrightarrow {z_p})$ \\
$d _p \gets \rm{maxDiff_1} ({[\theta _p},\overrightarrow{\theta^{all}_p}])$ \tcp{\footnotesize{searching the best position of the $p$-th particle}}
$p \gets p+1$
}
$d _{glo} \gets \rm{maxDiff_2} ([d_1,d_2,...,d_P])$ \tcp{\footnotesize{searching the best position among all particles}}

initialization $\overrightarrow {z_p^r} \gets \overrightarrow {z_p}$, $p \in [1,P]$ \\
\While{$r < R+1$}{
  \For{$p < P + 1$}{
     velocity and position updating as described in Eqn.~(\ref{eqn:pso1}) and (\ref{eqn:pso2}) \\
     best position searching as presented from Line 2 to 6 \\
     $p \gets p+1$
  }
  $r \gets r+1$
}
\KwResult{$Z$} 
\end{algorithm}

\subsection{Stealthy Perturbation}
\label{subsec:stealthy}
\begin{figure}[t]
\centering
\includegraphics[width=0.45\textwidth]{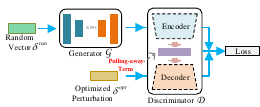}
\caption{Illustration of the EBGAN structure.}
\label{fig:ebgan}
\end{figure}

After considering the universality and robustness, \name needs to further explore the stealthiness of perturbation to avoid detection. Specifically, although our perturbations own the ability to complete cross-model attacks in the real-world setting, its single perturbation pattern is easily detected and then removed from the contaminated version with existing detection mechanisms. The static nature of one single perturbation greatly lower its stealthy ability. To escape this kind of detection, the most direct way is to generate multiple perturbations and then randomly utilize them as stated in~\cite{liu2023exploring} to present diverse attack patterns. However, this scheme is not feasible for \sname, because optimized perturbations do offer obviously distinct performance in real-world attack scenarios. Therefore, we choose the perturbation with the best attack performance among candidates in $Z$ and abandon the others. In our case, there is only one perturbation is available. To break this dilemma, we resort to the EBGAN model to generate multiple surrogates of one perturbation, thus ensuring the diversity of attack patterns. Moreover, compared to directly using multiple locally optimized perturbations, the surrogates of the best one can provide more significant perturbation capability (explained later). The structure of EBGAN utilized in \name is illustrated in Fig.~\ref{fig:ebgan}, which mainly consists of one generator $\mathcal{G}$ and one discriminator $\mathcal{D}$. The generator outputs the fake perturbation $\mathcal{G}(\delta^{ran})$ when inputting a random vector $\delta^{ran}$, while the discriminator aims to estimate the energy value distribution of both real and fake perturbations. EBGAN takes effect based on two criteria: $\mathcal{G}$ attempts to generate fake samples following a similar energy distribution to the real ones, thereby minimizing the sample reconstruction error in the en/de-coder and cheating $\mathcal{D}$; the discriminator aims to make the real samples reconstruction error small and make the fake one large. To construct this EBGAN model to meet the above two criteria, the objective function is defined as follows:
\begin{equation}\label{eqn:ganeqn1}
{L_{Dis}}(\delta^{opt},\delta^{ran}) = D(\delta^{opt}) + {[Thr - D(G(\delta^{ran}))]^ + }
\end{equation}
\begin{equation}\label{eqn:ganeqn2}
{L_{Gen}}(\delta^{ran}) = D(G(\delta^{ran}))
\end{equation}

\begin{figure}[b]
\subfigure[Sample distribution]{
\begin{minipage}[t]{0.46\linewidth}
\centering
\includegraphics[width=1\textwidth]{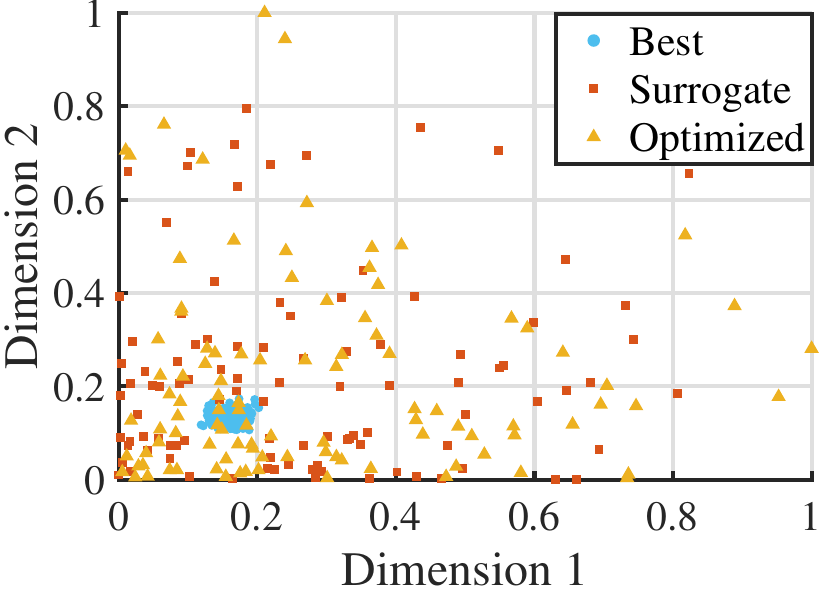}
\label{fig:distr1}
\end{minipage}
}
\subfigure[Feature difference]{
\begin{minipage}[t]{0.46\linewidth}
\centering
\includegraphics[width=1\textwidth]{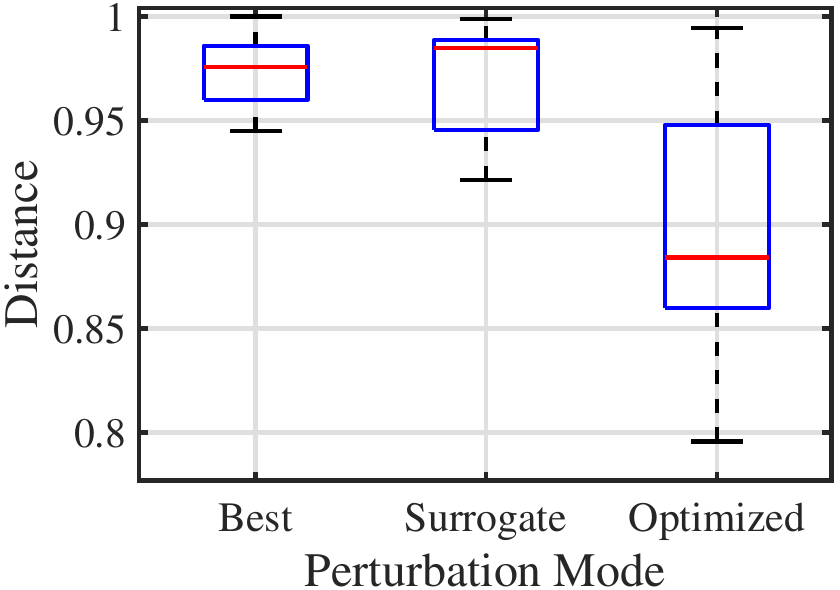}
\label{fig:diff2}
\end{minipage}
}
\caption{The distributions of the best perturbation, surrogates, and the optimized perturbations are shown in (a), while CSI amplitude and phase pattern differences are presented in (b).}
\label{fig:ebganPerf}
\end{figure}
where $[x]^+=\max\{0,x\}$ and ${[Thr - D(G(\delta^{ran}))]^ + }$ drives the reconstruction error of fake samples with a value higher than $Thr$. \name calculates the sample reconstruction error using the mean squared error. Moreover, to avoid overfitting and make the newly generated surrogate diverse, \name leverages pulling-away term~\cite{zhao2016energy} working on the encoder output layer to orthogonalize the pairwise sample representation. The generator and discriminator share the same 7-layer structure, i.e., $\{ \underbrace {\overbrace {N,128,32}^{Encoder},\mathop {16}\limits^{\scriptstyle~Embedding\hfill\atop \scriptstyle{\rm{   }}~~~~Layer\hfill} ,\overbrace {32,128,N}^{Decoder}}_{Generator/Discriminator}\}$. Subsequently, we evaluate the effectiveness of our surrogate generation mechanism with a preliminary experiment. On the one hand, Fig.~\ref{fig:distr1} shows the two-dimensional CSI distribution (compressed by t-SNE~\cite{vanizing}) when emitting the best perturbation one hundred times, one hundred surrogates, and the locally optimized perturbations. These surrogates enlarge the averaging sample distance of the best perturbation by more than ten times and keep a similar distribution with optimized ones. The result reveals that the surrogate greatly increases the attack pattern diversity. Moreover, we calculate the amplitude and phase feature differences in four sensing applications (as described in Sec.~\ref{subsec:feasibility}) under three cases: using the best perturbation, surrogates, and the optimized perturbations. As shown in Fig.~\ref{fig:diff2}, the feature difference distribution of surrogates is very similar to the best perturbation and much better than the last class. With this experimental result, we conclude that the proposed surrogate generation method not only ensures the perturbation performance but also improves the attack diversity.

\subsection{Putting All Things Together}
As illustrated in Fig.~\ref{fig:introScenario}, WiFi-based sensing systems leverage CSIs to present use states, thereby providing contactless services. In WiIntruder, an attacker launches one perturbation attack possessing three merits, namely, universality, robustness, and stealthiness, to impede normal services. It explores the importance differences of state-specific and model-dependent feature sub-components, and discriminatively treats them in the perturbation generation phase, to enhance the transferability of perturbations (cf. Sec.~\ref{subsec:uni}). Signal distortion caused by device desynchronization and wireless propagation are carefully handled in the perturbation optimization process as described in Sec.~\ref{subsec:roubust}. Finally, to ensure the stealthiness of attack, the EBGAN-based perturbation substitute generation mechanism is leveraged to improve the diversity of attack patterns.

\section{Evaluation}
\label{sec:evaluation}

\subsection{Experiment Setup}
\label{subsec:experimentSet}

\begin{figure}[t]
\centering
\includegraphics[width=0.45\textwidth]{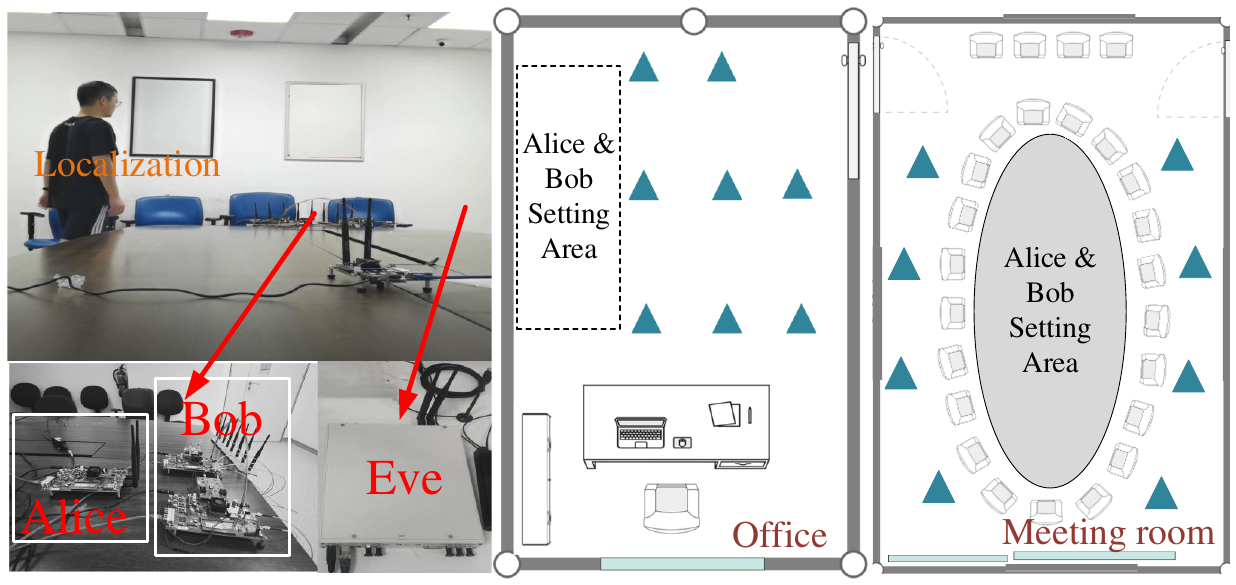}
\caption{Experiment settings and environment layouts: SDR devices are utilized to act as Alice, Bob, and Eve, respectively, for the indoor localization as an example.}
\label{fig:expSetting}
\end{figure}

\textbf{Hardware \& software.}
\name implementation relies on two widely-used SDR platforms as shown in Fig.~\ref{fig:expSetting}, i.e., WARP v3~\cite{WARP-web} and USRP X310~\cite{USRP-X310}. Each WARP node equipped with a FMC module Mango FMC-RF-2X245 enlarges the antenna quantity to four and the USRP uses one antenna by default. We leverage WARP nodes to act as Alice and Bob to complete legitimate sensing tasks, while the USRP plays Eve's role in emitting perturbation signals. Both WARP and USRP nodes are connected to a PC server through a 1 GHz Ethernet switch, which provides computation capability by an Intel Core i7-12700H CPU, 32GB RAM, and a GeForce RTX 2080 graphics card. The software framework implementation jointly employs GNU Radio, Pychram 2022, and MATLAB 2022b running on the server. Fig.~\ref{fig:expSetting} exhibits the experiment environment layout, i.e., a meeting room and a personal office. We respectively place Eve at eight different positions (marked as cyan triangles) to launch attacks.

\textbf{Utilized sensing models.}
We evaluate \sname's hazard using seven existing sensing models: Widar3~\cite{3326081} and C{\scriptsize ROSS}GR~\cite{li2021crossgr} for gesture recognition, SLNet~\cite{yang2023slnet} and ResMon~\cite{zheng2023resmon} for respiratory monitoring, MultiAuth~\cite{3467032} and Wi-PIGR~\cite{zhang2021wi} for user authentication, and DAFI~\cite{3494954} for indoor localization. We adjust experiment hardware configurations and parameters (e.g., antenna placement and packet transmission rate) to keep consistent with each sensing system's specific settings. For instance, Widar3 equips Alice with one antenna and Bob with six antennas, but Alice and Bob own three and one antenna, respectively, in C{\scriptsize ROSS}GR. In the following, we describe some key settings to help understand the process of reusing these models.

\textit{Gesture recognition.}
Widar3 totally recognizes six gestures, namely, \textit{push}, \textit{sweep}, \textit{clap}, \textit{slide}, \textit{draw circle}, and \textit{draw zigzag}. It makes the data and codes public\footnote{http://tns.thss.tsinghua.edu.cn/widar3.0/}, thus we can directly re-implement this system. Since Widar3's model is ready-made, each participant just needs to provide the test data by performing every gesture fifty times. In C{\scriptsize ROSS}GR, we reuse its convolutional neural network-based classifier to recognize fifteen gestures. The participant performs each gesture fifty times and we utilize its data augment module to enlarge the sample size tenfold. $Accuracy_1$ is utilized to measure the ratio of gesture CSI samples that are correctly recognized.

\textit{Respiratory monitoring.}
SLNet proposes a deep wireless sensing architecture to enhance spectrogram resolution to accurately monitor respiratory rate. Each participant here provides nineteen groups of CSI respiratory samples at different times and each group takes about three minutes. Furthermore, ResMon aims to detect the state (e.g., normal breathing and cough) of one target user relying on the basic respiratory sign. All participants provide the total number of two hundred CSI respiratory samples of each state. Breath per minute error between the estimated respiration rate and the ground truth is leveraged to measure the monitoring performance, and we leverage $Accuracy_2$ to present the correct ratio of each participant's respiratory state classification.

\textit{User authentication.}
Multi-Auth utilizes CSI spectrum patterns of specific user activities to represent individual identity. It authenticates users by distinguishing their common activities, including walking, sitting down, hand movement, etc. Each participant performs every activity one hundred times to provide CSI samples to register identity information. Wi-PIGR constructs the correspondence relationship between user identity and his/her gait information. Each participant walks along random paths for about twenty minutes and two-hundred gait instances are collected. We employ $Accuracy_3$ to assess the correct ratio of judging positive and false sample labels.

\textit{Indoor localization.}
DAFI attempts to address the inconsistency issue of CSI position fingerprints between the original and target domains. When the participant stays in one location, he/she is allowed to do any activity to fully present the CSI fingerprint traits. Each participant provides a total number of two thousand fingerprint samples collected at eight positions as shown in Fig.~\ref{fig:expSetting}. We leverage DAFI's neural network-based location classifier to judge one user's position. $Accuracy_4$ is the ratio between correct prediction positions and the total number of fingerprints.

\begin{figure*}[t]
\subfigure[C{\scriptsize ROSS}GR]{
\begin{minipage}[t]{0.23\linewidth}
\centering
\includegraphics[width=1\textwidth]{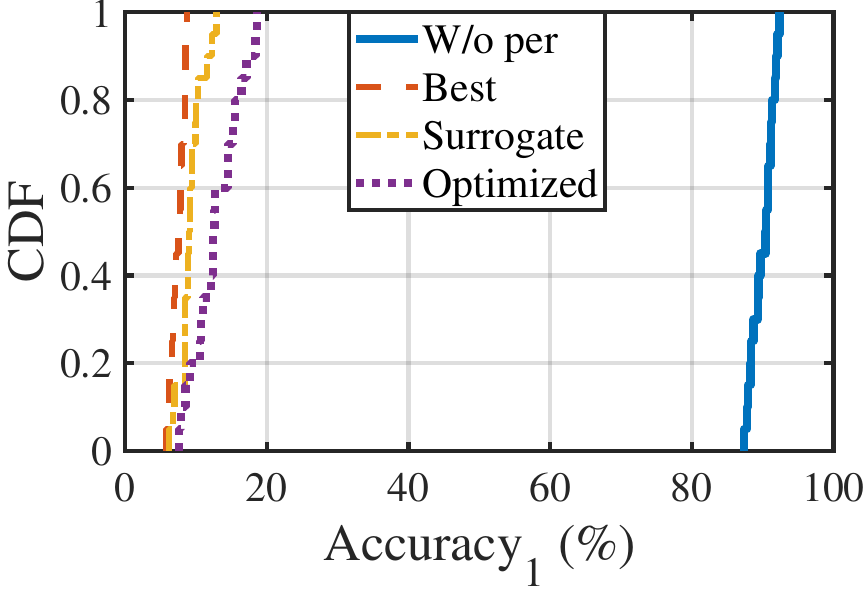}
\label{subfig:exp1crossgr}
\end{minipage}
}
\subfigure[ResMon]{
\begin{minipage}[t]{0.23\linewidth}
\centering
\includegraphics[width=1\textwidth]{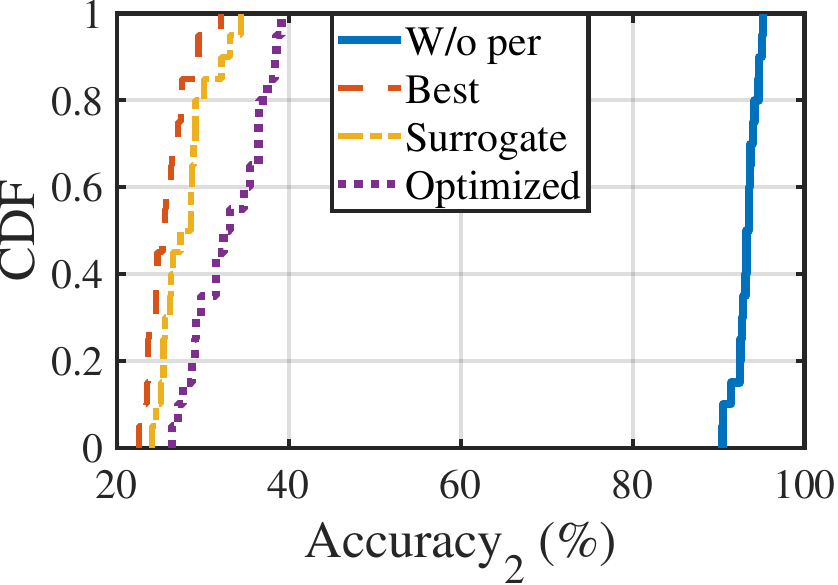}
\label{fig:exp1resmon}
\end{minipage}
}
\subfigure[Wi-PIGR]{
\begin{minipage}[t]{0.23\linewidth}
\centering
\includegraphics[width=1\textwidth]{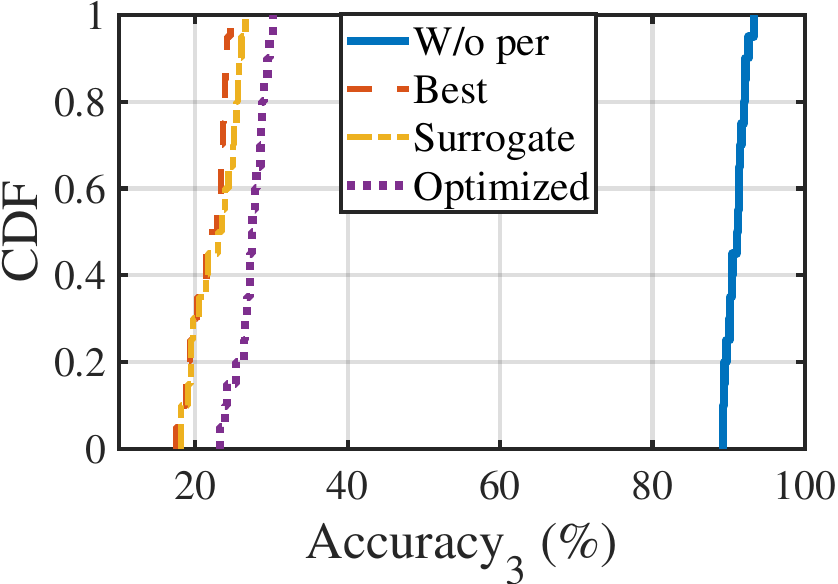}
\label{subfig:exp1wipigr}
\end{minipage}
}
\subfigure[DAFI]{
\begin{minipage}[t]{0.23\linewidth}
\centering
\includegraphics[width=1\textwidth]{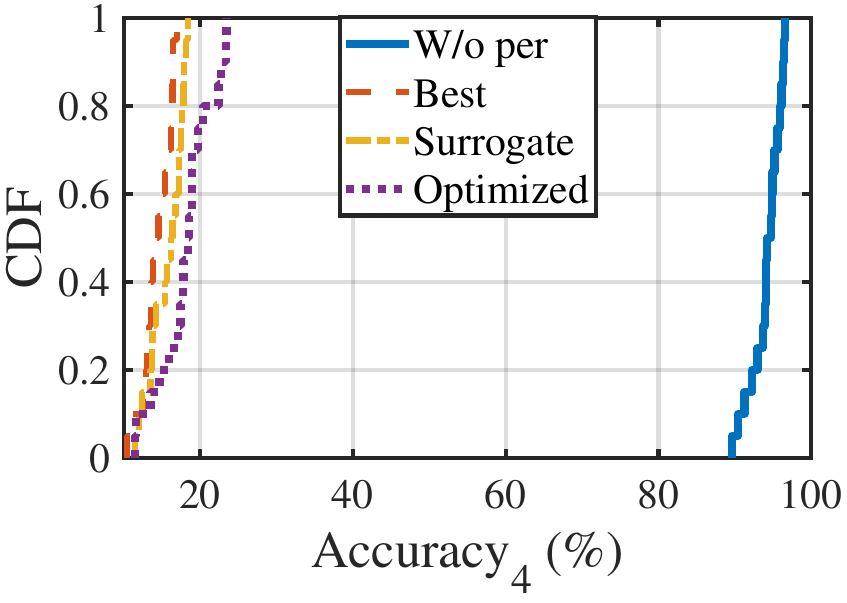}
\label{subfig:exp1dafi}
\end{minipage}
}
\caption{Performance of sensing applications before and after enabling perturbation: (a) gesture recognition, (b) respiratory state monitoring, (c) user authentication, and (d) indoor localization.}
\label{fig:overall}
\end{figure*}

\textit{Notes.}
Three models (i.e., Widar3, SLNet, and MultiAuth) act as the white-box role to help us construct the perturbation generation mechanism, while the remaining models are \sname's attack targets. What is noteworthy is that the model and data of the localization application do not involve perturbation optimization. The goal of this setting is to evaluate \sname's performance in an ``unseen" application, which meets the black-box manner. We employ the same ratio of the collected samples as the reference work to train and the rest for testing. Data collection steps are conducted respectively in two environments. Except for the above content, other settings remain consistent with the reference works.

\textbf{Perturbation generation.}
We train Widar3, SLNet, and MultiAuth relying on their original rules and record gradients $\overline{\Delta^u}$. Subsequently, the PSO method described in Sec.~\ref{subsec:signalGen} generates a near-optimal perturbation set $Z$ consisting of one hundred samples and the optimization process stops until the iteration epoch reaches one thousand. All samples in $Z$ are labeled as 1, acting as the true class. For each perturbation, we randomly change the element of each perturbation in the range of $[-\varepsilon, \varepsilon]$ to generate ${\delta ^{ran}}$. $\varepsilon$ is set to the maximal energy variation of the user motion existing period and $Thr$ is set to $\varepsilon$. We train the generator and the discriminator in turn, fixing the untrained one's parameters accordingly. The random parameter dropout ratio is set to 0.5 and the batch size is 64. The iteration termination condition is the averaging loss less than 0.1. The Adam optimizer is utilized to update parameters, with a learning rate of 0.01. 

\subsection{Evaluation}
\textbf{Overall performance.}
\label{subsec:overallPer}
The primary task is to evaluate the performance of four sensing applications when using the best perturbation, its surrogates, and the optimized ones, respectively. In this experiment, we launch the three-type perturbations (including three thousand CSI samples), respectively, at ten different times and record the corresponding performance. As shown in Fig.~\ref{fig:overall}, the best perturbation and surrogates lead to similar performance dropping, with the averaging differences of four metrics as 1.8\%, 2.1\%, 0.8\%, and 1.3\%. The results are consistent with our previous analysis that surrogates maintain competitive aggressiveness. Moreover, surrogates result in an obvious performance decrease in all applications, which are 79.6\%, 65.2\%, 68.3\%, and 78.4\%. Specifically, the accuracy of recognizing fifteen gestures decreases to 9.2\%, close to the rate of a blind guess among these categories; user authentication accuracy drops to 22.5\%, thus the verification system incorrectly judges user identity with high probability. The other two applications exhibit similar performance variation. To sum up, our proposed perturbation mechanism imposes a considerable security threat against WiFi-based wireless sensing systems in such a real-world setting, which should be highly concerned.

\begin{figure}[t]
\centering
\includegraphics[width=0.45\textwidth]{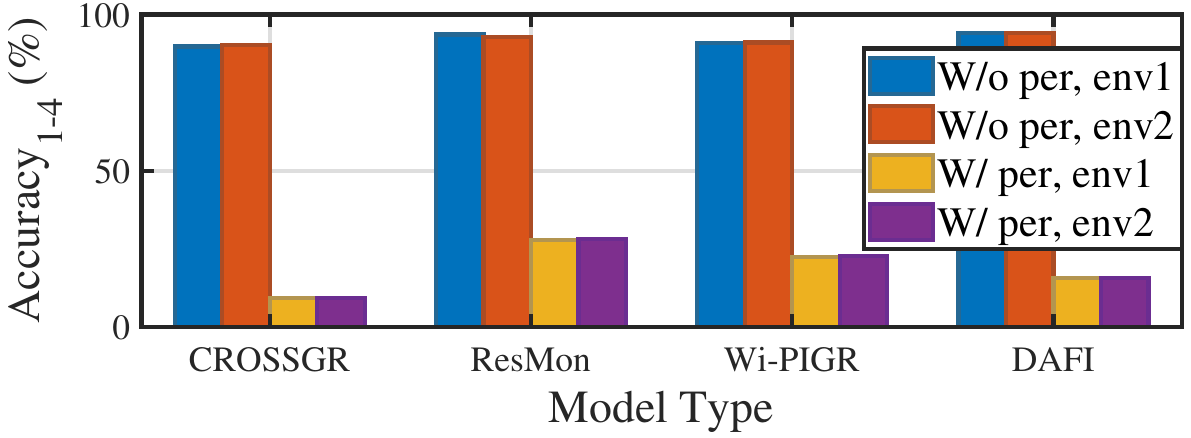}
\caption{Performance of four sensing applications when turning on and off perturbation in the meeting room (env1) and office (env2).}
\label{fig:env}
\end{figure}

\textbf{Impact of environment differences}.
Wireless channel states across environments are distinct and have important impacts on perturbation performance. To further explore the attack's practicality, we study whether the proposed mechanism maintains consistent aggressiveness in different environments. As mentioned in the previous experiment setting, we conducted the same data collection and model training operations in our meeting room and office. The results are illustrated in Fig.~\ref{fig:env}, which shows the performance of four applications in the two environments. There are two key observations as follows: without perturbing, legitimate users obtain similar high-quantity service in two environments, which evidences that we have successfully reused these models; the performance dropping in all applications is consistent with both environments, with differences of 0.5\%, 0.7\%, 0.2\%, and 0.1\%, respectively. These results clarify that \name maintains outstanding performance across environments, which further confirms its practicality.

\begin{figure}[h]
\begin{minipage}[t]{0.48\linewidth}
\centering
\includegraphics[width=1\textwidth]{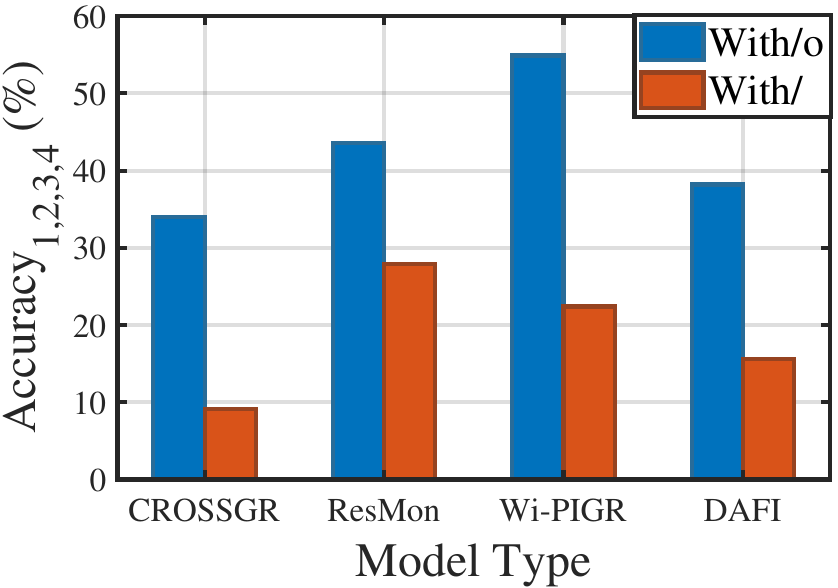}
\caption{Performance with and without the use of the importance weighting mechanism.}
\label{fig:trans}
\end{minipage}
\hspace{0.1em}
\begin{minipage}[t]{0.48\linewidth}
\centering
\includegraphics[width=1\textwidth]{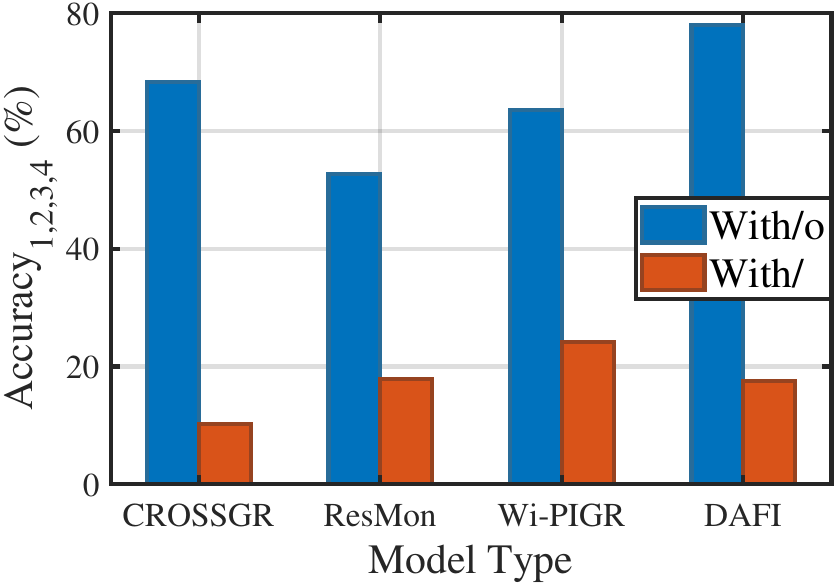}
\caption{Performance variation when considering the device and time desynchronization interference.}
\label{fig:robust}
\end{minipage}
\end{figure}

\textbf{Impact of importance weight mechanism.}
Importance weight-driven user state-specific feature difference maximization is proposed in Sec.~\ref{subsec:uni} to enhance the transferability and hence the universality of perturbation. In this part, we study the effectiveness of this mechanism, by comparing the performance variations with and without enabling it. In each setting, we collect one thousand contaminated CSI samples from four sensing applications. As presented in Fig.~\ref{fig:trans}, the importance weight mechanism brings an obvious performance dropping in all applications, that are 24.9\%, 15.7\%, 32.5\%, and 22.6\%. These results indicate that the mechanism is effective in enhancing the transferability of perturbation destructiveness across distinct models.

\textbf{Impact of time and device desynchronization.}
We detail the impact of time and device desynchronization on perturbation attacks via a perspective of theoretical analysis in Sec.~\ref{subsec:roubust}, while removing their interference in the optimization process. This experiment is conducted to verify whether the module contributes to the attack's practicality in a real-world setting. We collect one thousand contaminated CSI samples with and without considering this desynchronization factor. Subsequently, Fig.~\ref{fig:robust} presents the performance variation when suspending this module. The average accuracy of four sensing applications is reduced by 58.1\%, 34.8\%, 39.5\%, and 60.4\%. From these results, we can conclude that the proposed module is effective and critical for the attack implementation.

\textbf{Impact of a single perturbation duration.}
Survival duration (switching frequency) of one perturbation surrogate directly affects the attack pattern in the temporal dimension, which may change the attack performance. To study its impact, we adjust the value from 0.02~\!s to 8~\!s with a step size of 0.02~\!s (in the range from 0.02 to 0.2~\!s) and 0.2~\!s, respectively, while recording the corresponding classification accuracy. As shown in Fig.~\ref{fig:duration}, as survival duration (in the range beyond 0.2~\!s) increases, the accuracy variation first shows a slowly increasing trend and then fluctuates within a narrow range in all models. This variation trend is reasonable, reflected in two aspects. First, switching perturbation frequently means that the local CSI trait of one user activity can be adequately destroyed. Therefore, improving the switching frequency leads to accuracy decreasing. Second, when the survival duration is relatively large, the accuracy remains basically unchanged, because the impact imposed on local features of CSIs is limited. Moreover, when the duration is less than 0.2~\!s, the accuracy is stable. Therefore, fully considering the performance, \name lets Eve switch the perturbation every 0.2~\!s by default.

\subsection{Countermeasures}
\begin{figure}[h]
\begin{minipage}[t]{0.48\linewidth}
\centering
\includegraphics[width=1\textwidth]{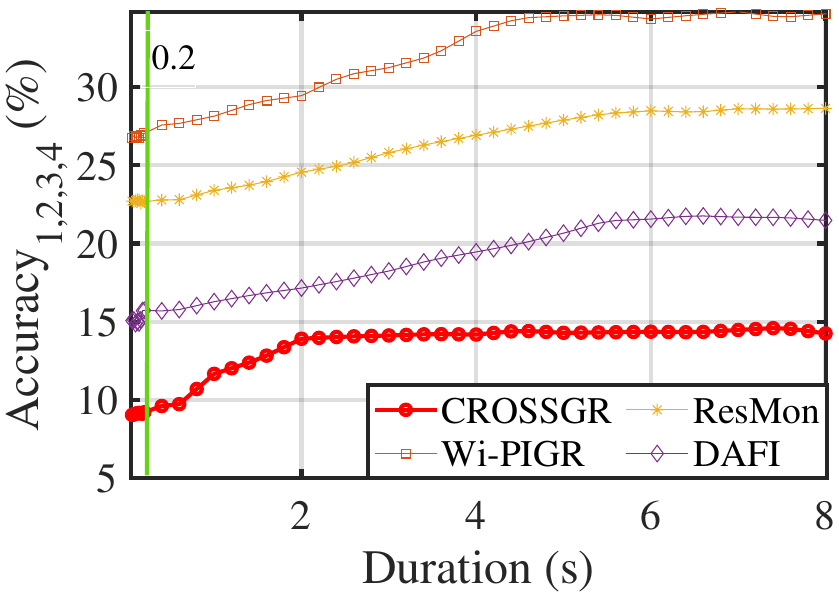}
\caption{Performance variation when adjusting the duration from 0.2~\!s to 8~\!s.}
\label{fig:duration}
\end{minipage}
\hspace{0.1em}
\begin{minipage}[t]{0.48\linewidth}
\centering
\includegraphics[width=1\textwidth]{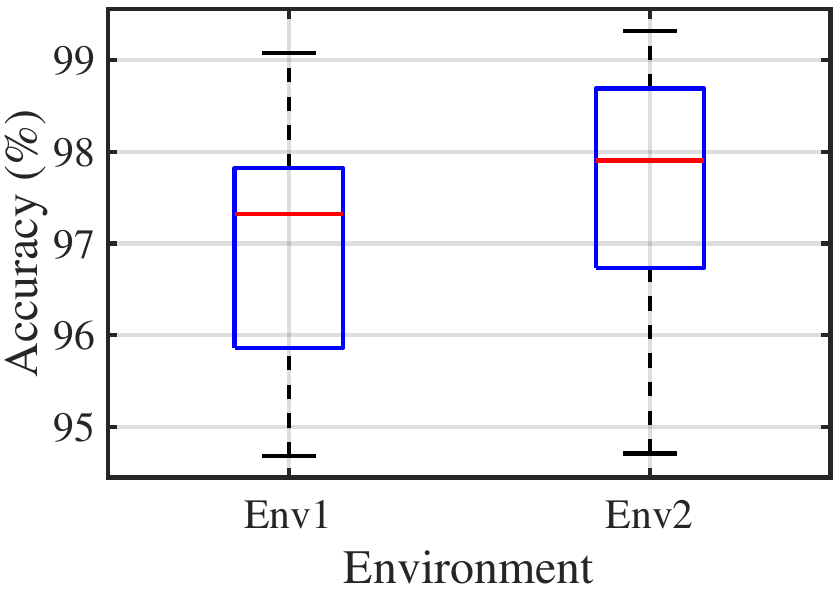}
\caption{Detection performance under two experiment environments.}
\label{fig:defensing}
\end{minipage}
\end{figure}

After fully learning \sname's hazard, the following task is to design effective detection strategies against it. Inspired by existing defense mechanisms~\cite{3465397}, we realize that endowing sensing models with the ability to recognize perturbations by adversarial training operation is a useful protection way. To be specific, CSI perturbation samples should be treated as one class of training data and labeled as 0 in our case. During the model training phase, we process perturbations leveraging the same procedure as the legitimate data. We collect the two-type (i.e., ${H_k}+{H_{Eve}}{H_\delta}$ and ${H_{Eve}}{H_\delta}$) perturbations in the following few cases, which are turning on and off legal sensing services when Eve emits perturbations. For each perturbation type and environment setting, we collect five hundred samples for every sensing application. We retrain four sensing models as the original works and then using them respectively recognize the perturbation samples. As shown in Fig.~\ref{fig:defensing}, the averaging detection accuracy in two environments are both larger than 97\%. Although this method increases the time cost by about forty minutes during the model training phase, it provides high accuracy for adversarial perturbation detection, which greatly helps sensing systems effectively defend against \sname. 

Moreover, we consider the fundamental reason for successfully launching perturbation attacks from the perspective of wireless communication protocol. In IEEE 802.11, LTS is
a public field in OFDM preambles and all devices (including Eve) know it. Therefore, Eve can monitor the LTS field without any restrictions, further decoding $H_{Eve}$ between it and legal devices. $H_{Eve}$ is critical for practical perturbation generation. Based on the above analysis, keeping LTS confidential or dynamically changing it can effectively prevent Eve from inferring channel state information and further launching attacks. Although this approach sounds simple, its implementation is challenging,  because of significant impacts on existing communication services. For example, unauthorized devices cannot decode authentic CSIs and hence data from AP beacons, thus will be denied to join new wireless networks. Besides, this implies that the WiFi standards have to be revised as well, which poses more challenges. Nevertheless, the exposure of security issues on existing protocols motivates us to reexamine their security, which will impose a positive impact on future protocol design and development. 

\section{Discussions and Future Works}
\label{sec:discussion}

\textit{Antenna quantity.}
Eve's aggressivity and its antenna quantity have a positive impact in the sense that increasing the quantity improves the perturbation pattern diversity in both digital and physical spaces. On the one hand, if Eve is equipped with more antennas, the candidate perturbation set accordingly expands, which means that \name can search over a larger digital space for generating the near-optimal perturbation and thereby launching more powerful attacks. On the other hand, Eve with multiple antennas leveraging MIMO beamforming technology~\cite{ding2015energy} can also precode perturbations before transmitting them, which provides flexible control over the perturbation power distribution and direction in physical space. Antenna quantity determines the control ability of the space granularity and perturbation pattern to design. In our attack scenario settings, to ensure Eve's physical stealthiness, we assume that it owns only one antenna. Therefore, a comprehensive study on the relationship between antenna quantity and Eve's aggressivity is intriguing and will be investigated in our future work.

\textit{Multi-Bob scenario.}
\name currently verifies the effectiveness of launching practical perturbation attacks in the one-user (Bob) scenario. Nevertheless, the existence of multiple Bobs deserves further exploration. One may wonder whether the experiment in Sec.~\ref{subsec:overallPer} respectively places one Bob at distinct eight positions (relative to Eve) is equivalent to enabling multiple Bobs simultaneously. The answer is no. Specifically, common WiFi APs are configured with MIMO beamforming aiming to provide multiple coexisting users with high-quality data transmission services. This means that Alice dynamically adjusts the precoding mode based on each Bob's wireless channel, which is the main difference compared with the ``static'' channel adjustment as only one Bob exists. To sum up, our work focusing on one Bob scenario can be regarded as the footstone to expand this attack to multi-user settings.

\textit{Other sensing medium.}
Except for WiFi, other common mediums such as Radar and ultrasound have also been leveraged to build innovative wireless sensing systems. Deep learning technology exhibits strong penetration in these sensing communities, which requires them to properly handle the threats from perturbation attacks. In such a ubiquitous sensing environment, it is of paramount importance to understand the performance of these applications in face of perturbation attacks. Unfortunately, the research results of \name cannot directly be adapted to other sensing applications due to the differences in their operational mechanisms. However, the analysis techniques for \name provide hints to study the practicality of attacks, especially some issues arising in real-world attacks, such as signal distortion and device desynchronization. Overall, investigation on the threats of perturbation attacks to common sensing applications and the corresponding defense strategies is important for securing sensing systems.

\section{Conclusion}
\label{sec:conclusion}
In this paper, we have explored practical perturbation attack named \name to fully examine its hazard on WiFi-based sensing systems, while pointing out potential defense mechanisms. Through careful examination of the internal distinction across models, we have proposed much more general and effective attacks on WiFi-based sensing systems, where an importance weight-driven feature difference maximization mechanism has been utilized to enable the attack universality. To ensure robustness, we have carefully studied the impacts of device desynchronization and wireless propagation distortion imposed on perturbation signals and handled them in the optimization formulation. To tackle attack stealthiness, we have leveraged the optimal perturbation surrogates provided by an energy-based generative adversarial network and generate random hopping patterns to enhance the diversity of attack patterns. Extensive experiments have demonstrated the outstanding performance of \name in compromising four common WiFi-based sensing services. We hope through the comprehensive analysis of the hazard of perturbation attacks, we can develop effective countermeasures and better protect WiFi-based sensing systems.

\bibliographystyle{IEEEtran}
\bibliography{reference}

\begin{thebibliography}{10}
\providecommand{\url}[1]{#1}
\csname url@samestyle\endcsname
\providecommand{\newblock}{\relax}
\providecommand{\bibinfo}[2]{#2}
\providecommand{\BIBentrySTDinterwordspacing}{\spaceskip=0pt\relax}
\providecommand{\BIBentryALTinterwordstretchfactor}{4}
\providecommand{\BIBentryALTinterwordspacing}{\spaceskip=\fontdimen2\font plus
\BIBentryALTinterwordstretchfactor\fontdimen3\font minus
  \fontdimen4\font\relax}
\providecommand{\BIBforeignlanguage}[2]{{%
\expandafter\ifx\csname l@#1\endcsname\relax
\typeout{** WARNING: IEEEtran.bst: No hyphenation pattern has been}%
\typeout{** loaded for the language `#1'. Using the pattern for}%
\typeout{** the default language instead.}%
\else
\language=\csname l@#1\endcsname
\fi
#2}}
\providecommand{\BIBdecl}{\relax}
\BIBdecl

\bibitem{ma2019wifi}
Y.~Ma, G.~Zhou, and S.~Wang, ``{WiFi Sensing with Channel State Information: A
  Survey},'' \emph{ACM Computing Surveys}, vol.~52, no.~3, pp. 1--36, 2019.

\bibitem{9796740}
H.~Kong, L.~Lu, J.~Yu, and et~al, ``{Push the Limit of WiFi-based User
  Authentication towards Undefined Gestures},'' in \emph{IEEE INFOCOM},
  Virtual, 2022.

\bibitem{8613849}
Y.~Meng, J.~Li, H.~Zhu, and et~al, ``{Revealing Your Mobile Password via WiFi
  Signals: Attacks and Countermeasures},'' \emph{IEEE Transactions on Mobile
  Computing}, vol.~19, no.~2, pp. 432--449, 2020.

\bibitem{3485936}
R.~Xiao, J.~Liu, J.~Han, and et~al, ``{OneFi: One-Shot Recognition for Unseen
  Gesture via COTS WiFi},'' in \emph{ACM SenSys}, Coimbra, Portugal, 2021.

\bibitem{9141400}
C.~Li, M.~Liu, and Z.~Cao, ``{WiHF: Gesture and User Recognition With WiFi},''
  \emph{IEEE Transactions on Mobile Computing}, vol.~21, no.~2, pp. 757--768,
  2022.

\bibitem{wang2016device}
J.~Wang, X.~Zhang, Q.~Gao, H.~Yue, and H.~Wang, ``Device-free wireless
  localization and activity recognition: A deep learning approach,'' \emph{IEEE
  Transactions on Vehicular Technology}, vol.~66, no.~7, pp. 6258--6267, 2016.

\bibitem{wang2020learning}
J.~Wang, Q.~Gao, X.~Ma, Y.~Zhao, and Y.~Fang, ``{Learning to Sense: Deep
  Learning for Wireless Sensing with Less Training Efforts},'' \emph{IEEE
  Wireless Communications}, vol.~27, no.~3, pp. 156--162, 2020.

\bibitem{wang2018device}
J.~Wang, Q.~Gao, M.~Pan, and Y.~Fang, ``Device-free wireless sensing:
  Challenges, opportunities, and applications,'' \emph{IEEE Network}, vol.~32,
  no.~2, pp. 132--137, 2018.

\bibitem{3326081}
Y.~Zheng, Y.~Zhang, K.~Qian, and et~al, ``{Zero-Effort Cross-Domain Gesture
  Recognition with Wi-Fi},'' in \emph{ACM MobiSys}, Seoul, South Korea, 2019.

\bibitem{3467032}
H.~Kong, L.~Lu, J.~Yu, and et~al, ``{MultiAuth: Enable Multi-User
  Authentication with Single Commodity WiFi Device},'' in \emph{ACM MobiHoc},
  Shanghai, China, 2021, pp. 31--40.

\bibitem{xiong2013arr}
J.~Xiong and K.~Jamieson, ``{Arraytrack: A Fine-grained Indoor Location
  System},'' in \emph{Lombard, IL}, Atlanta, USA \& Cambridge, UK, 2013.

\bibitem{3241548}
W.~Jiang, C.~Miao, F.~Ma, and et~al, ``{Towards Environment Independent Device
  Free Human Activity Recognition},'' in \emph{ACM MobiCom}, New Delhi, India,
  2018.

\bibitem{3423348}
Z.~Li, Y.~Wu, J.~Liu, and et~al, ``{AdvPulse: Universal, Synchronization-Free,
  and Targeted Audio Adversarial Attacks via Subsecond Perturbations},'' in
  \emph{ACM CCS}, Virtual, 2020.

\bibitem{chc1}
H.~Cao, H.~Jiang, D.~Liu, and J.~Xiong, ``Evidence in hand: Passive vibration
  response-based continuous user authentication,'' in \emph{2021 IEEE 41st
  International Conference on Distributed Computing Systems (ICDCS)}, 2021, pp.
  1020--1030.

\bibitem{LiNPSKRS19}
S.~Li, A.~Neupane, S.~Paul, and et~al, ``{Stealthy Adversarial Perturbations
  Against Real-Time Video Classification Systems},'' in \emph{NDSS}, San Diego,
  California, 2019.

\bibitem{nan2023you}
Y.~Nan, X.~Wang, L.~Xing, X.~Liao, and et~al, ``{Are You Spying on Me?
  Large-Scale Analysis on IoT Data Exposure through Companion Apps},'' in
  \emph{USENIX Security Symposium}, Anaheim, USA, 2023.

\bibitem{3534618}
Y.~Zhou, H.~Chen, C.~Huang, and et~al, ``{WiAdv: Practical and Robust
  Adversarial Attack against WiFi-Based Gesture Recognition System},''
  \emph{ACM UbiComp/IMWUT}, 2019.

\bibitem{liu2023exploring}
Z.~Liu, C.~Xu, E.~Sie, and et~al, ``{Exploring Practical Vulnerabilities of
  Machine Learning-based Wireless Systems},'' in \emph{USENIX NSDI}, Boston,
  USA, 2023.

\bibitem{9796920}
J.~Liu, Y.~He, C.~Xiao, and et~al, ``{Physical-World Attack towards WiFi-based
  Behavior Recognition},'' in \emph{IEEE INFOCOM}, Virtual, 2022.

\bibitem{3484777}
A.~Bahramali, M.~Nasr, A.~Houmansadr, and et~al, ``{Robust Adversarial Attacks
  Against DNN-Based Wireless Communication Systems},'' in \emph{ACM CCS},
  Virtual, 2022.

\bibitem{9609969}
B.~Kim, Y.~E. Sagduyu, K.~Davaslioglu, and et~al, ``{Channel-Aware Adversarial
  Attacks Against Deep Learning-Based Wireless Signal Classifiers},''
  \emph{IEEE Transactions on Wireless Communications}, vol.~21, no.~6, pp.
  3868--3880, 2022.

\bibitem{8792120}
B.~Flowers, R.~M. Buehrer, and W.~C. Headley, ``{Evaluating Adversarial Evasion
  Attacks in the Context of Wireless Communications},'' \emph{IEEE Transactions
  on Information Forensics and Security}, vol.~15, no.~1, pp. 1102--1113, 2020.

\bibitem{yingyingchen}
Y.~Xie, R.~Jiang, X.~Guo, and et~al, ``{Universal Targeted Adversarial Attacks
  Against mmWave-based Human Activity Recognition},'' in \emph{IEEE INFOCOM},
  New York area, USA, 2023.

\bibitem{ZhangZL0CZH21}
W.~Zhang, S.~Zhao, L.~Liu, and et~al, ``{Attack on Practical Speaker
  Verification System Using Universal Adversarial Perturbations},'' in
  \emph{IEEE ICASSP}, Toronto, Ontario, Canada, 2021.

\bibitem{WangHCLCW22}
R.~Wang, Z.~Huang, Z.~Chen, and et~al, ``{Anti-Forgery: Towards a Stealthy and
  Robust DeepFake Disruption Attack via Adversarial Perceptual-aware
  Perturbations},'' in \emph{IJCAI}, Messe Wien, Austria, 2022.

\bibitem{XieWKH22}
S.~Xie, H.~Wang, Y.~Kong, and et~al, ``Universal 3-dimensional perturbations
  for black-box attacks on video recognition systems,'' in \emph{IEEE S\&P},
  San Francisco, CA, 2022.

\bibitem{3465397}
X.~Zhang, X.~Zheng, and W.~Mao, ``{Adversarial Perturbation Defense on Deep
  Neural Networks},'' \emph{ACM Computing Surveys}, vol.~54, no.~8, pp. 1--36,
  2021.

\bibitem{iu2014chann}
Y.~Liu, Z.~Tan, H.~Hu, and et~al, ``{Channel Estimation for OFDM},'' \emph{IEEE
  Communications Surveys \& Tutorials}, vol.~16, no.~4, pp. 1891--1908, 2014.

\bibitem{572909}
H.~Cao, D.~Liu, H.~Jiang, R.~Wang, Z.~Chen, and J.~Xiong, ``Lipauth:
  Hand-dependent light intensity patterns for resilient user authentication,''
  \emph{ACM Trans. Sen. Netw.}, 2023.

\bibitem{6847948}
C.~Han, K.~Wu, Y.~Wang, and et~al., ``{WiFall: Device-free Fall Detection by
  Wireless Networks},'' Toronto, Canada, 2014.

\bibitem{chc}
J.~Luo, H.~Cao, H.~Jiang, Y.~Yang, and C.~Zhe, ``mimocrypt: Multi-user
  privacy-preserving wi-fi sensing via mimo encryption,'' in \emph{2024 IEEE
  Symposium on Security and Privacy (SP)}, 2024.

\bibitem{9941045}
Z.~Chen, T.~Zheng, C.~Hu, and et~al, ``{ISACoT: Integrating Sensing with Data
  Traffic for Ubiquitous IoT Devices},'' \emph{IEEE Communications Magazine},
  vol.~61, no.~5, pp. 98--104, 2023.

\bibitem{3411816}
Y.~Zeng, D.~Wu, J.~Xiong, and et~al, ``{MultiSense: Enabling Multi-Person
  Respiration Sensing with Commodity WiFi},'' \emph{ACM UbiComp/IMWUT}, 2020.

\bibitem{9380161}
B.~Huang, R.~Yang, B.~Jia, and et~al, ``{A Theoretical Analysis on Sampling
  Size in WiFi Fingerprint-Based Localization},'' \emph{IEEE TVT}, vol.~70,
  no.~4, 2021.

\bibitem{3534574}
B.~Guo, W.~Zuo, S.~Wang, and et~al, ``{WePos: Weak-Supervised Indoor
  Positioning with Unlabeled WiFi for On-Demand Delivery},'' \emph{ACM
  UbiComp/IMWUT 2022}, 2022.

\bibitem{huang2021wars}
P.~Huang, X.~Zhang, S.~Yu, and et~al, ``{IS-WARS: Intelligent and Stealthy
  Adversarial Attack to Wi-Fi-based Human Activity Recognition Systems},''
  \emph{IEEE Transactions on Dependable and Secure Computing}, vol.~19, no.~6,
  pp. 3899--3912, 2021.

\bibitem{ChengDPSZ19}
S.~Cheng, Y.~Dong, T.~Pang, and et~al, ``Improving black-box adversarial
  attacks with a transfer-based prior,'' in \emph{NeurIPS}, Virtual, 2019.

\bibitem{yang2020adversarial}
J.~Yang, R.~Xu, R.~Li, and et~al, ``An adversarial perturbation oriented domain
  adaptation approach for semantic segmentation,'' in \emph{AAAI}, New York,
  USA, 2017.

\bibitem{li2021playing}
X.~Li, Y.~Jiang, C.~Liu, S.~Liu, H.~Luo, and S.~Yin, ``{Playing against
  Deep-Neural-Network-Based Object Detectors: A Novel Bidirectional Adversarial
  Attack Approach},'' \emph{IEEE Transactions on Artificial Intelligence},
  vol.~3, no.~1, pp. 20--28, 2021.

\bibitem{wei2020heuristic}
Z.~Wei, J.~Chen, X.~Wei, and et~al, ``{Heuristic Black-box Adversarial Attacks
  on Video Recognition Models},'' in \emph{AAAI}, New York, USA, 2020.

\bibitem{zhaneprint}
L.~Zhang, Y.~Meng, J.~Yu, and et~al, ``{Voiceprint Mimicry Attack Towards
  Speaker Verification System in Smart Home},'' in \emph{IEEE INFOCOM},
  Virtual, 2020, pp. 377--386.

\bibitem{cimysis}
L.~Cimini, ``Analysis and simulation of a digital mobile channel using
  orthogonal frequency division multiplexing,'' \emph{IEEE Transactions on
  Computers}, vol.~33, no.~7, pp. 665--675, 1985.

\bibitem{yang2023slnet}
Z.~Yang, Y.~Zhang, K.~Qian, and et~al, ``{SLNet: A Spectrogram Learning Neural
  Network for Deep Wireless Sensing},'' in \emph{USENIX NSDI}, Boston, USA,
  2023.

\bibitem{3494954}
H.~Li, X.~Chen, J.~Wang, and et~al, ``{DAFI: WiFi-Based Device-Free Indoor
  Localization via Domain Adaptation},'' \emph{ACM UbiComp/IMWUT}, 2022.

\bibitem{zhao2016energy}
J.~Zhao, M.~Mathieu, and Y.~LeCun, ``{Energy-based Generative Adversarial
  Network},'' in \emph{ICLR}, Toulon, France, 2017.

\bibitem{salzmann2021learning}
M.~Salzmann \emph{et~al.}, ``Learning transferable adversarial perturbations,''
  \emph{NeurIPS}, 2021.

\bibitem{huang2019enhancing}
Q.~Huang, I.~Katsman, H.~He, and et~al, ``Enhancing adversarial example
  transferability with an intermediate level attack,'' in \emph{IEEE/CVF ICCV},
  Seoul, Korea, 2019, pp. 4733--4742.

\bibitem{wang2021feature}
Z.~Wang, H.~Guo, Z.~Zhang, and et~al, ``Feature importance-aware transferable
  adversarial attacks,'' in \emph{IEEE/CVF ICCV}, Virtual, 2020.

\bibitem{nagantial}
S.~Nagaraj, S.~Khan, C.~Schlegel, and et~al, ``Differential preamble detection
  in packet-based wireless networks,'' \emph{IEEE Transactions on Wireless
  Communications}, vol.~8, no.~2, pp. 599--607, 2009.

\bibitem{zhu2017calibrating}
J.~Zhu, Y.~Im, S.~Mishra, and et~al, ``Calibrating time-variant,
  device-specific phase noise for cots wifi devices,'' in \emph{ACM Sensys},
  Delft, The Netherlands, 2020.

\bibitem{kotaru2015spotfi}
M.~Kotaru, K.~Joshi, D.~Bharadia, and et~al, ``Spotfi: Decimeter level
  localization using wifi,'' in \emph{ACM SIGCOMM}, London, United Kingdom,
  2015.

\bibitem{ZengL0LW021}
Y.~Zeng, J.~Liu, J.~Xiong, and et~al, ``Exploring multiple antennas for
  long-range wifi sensing,'' \emph{ACM IMWUT/UbiComp}, 2021.

\bibitem{9673102}
J.~Choi, ``{Sensor-Aided Learning for Wi-Fi Positioning With Beacon Channel
  State Information},'' \emph{IEEE Transactions on Wireless Communications},
  vol.~21, no.~7, pp. 5251--5264, 2022.

\bibitem{80211}
{Wikimedia}, ``{IEEE 802.11},''
  \url{https://en.wikipedia.org/wiki/IEEE_802.11}, 2023, online; accessed 13
  Dctober 2023.

\bibitem{ebericle}
R.~Eberhart and J.~Kennedy, ``{Particle Swarm Optimization},'' in \emph{IEEE
  ICNN}, vol.~4, 1995, pp. 1942--1948.

\bibitem{vanizing}
H.~Cao, H.~Jiang, D.~Liu, and et~al, ``{Evidence in Hand: Passive Vibration
  Response-based Continuous User Authentication},'' 2021.

\bibitem{WARP-web}
{Mango Communications}, ``{WARP v3 Kit},''
  \url{http://mangocomm.com/products/kits/warp-v3-kit/}, 2023, online; accessed
  10 October 2023.

\bibitem{USRP-X310}
{NI}, ``{USRP X310},'' \url{https://www.ettus.com/all-products/x310-kit/},
  2023, online; accessed 10 October 2023.

\bibitem{li2021crossgr}
X.~Li, L.~Chang, F.~Song, and et~al, ``{CrossGR: Accurate and Low-cost
  Cross-target Gesture Recognition Using Wi-Fi},'' \emph{ACM UbiComp/IMWUT},
  2021.

\bibitem{zheng2023resmon}
L.~Zheng, S.~Bi, S.~Wang, and et~al, ``{ResMon: Domain-adaptive Wireless
  Respiration State Monitoring via Few-shot Bayesian Deep Learning},''
  \emph{IEEE Internet of Things Journal}, 2023.

\bibitem{zhang2021wi}
L.~Zhang, C.~Wang, and D.~Zhang, ``{Wi-PIGR: Path Independent Gait Recognition
  with Commodity Wi-Fi},'' \emph{IEEE Transactions on Mobile Computing},
  vol.~21, no.~9, pp. 3414--3427, 2021.

\bibitem{ding2015energy}
H.~Ding, H.~Yue, J.~Liu, P.~Si, and Y.~Fang, ``{Energy-efficient Secondary
  Traffic Scheduling with MIMO Beamforming},'' in \emph{IEEE GLOBECOM}, San
  Diego, USA, 2015.

\end{thebibliography}

\end{document}